\documentclass[pre,twocolumn,showpacs,showkeys,superscriptaddress,preprintnumbers,floatfix]{revtex4}

\usepackage{url}
\usepackage{graphicx}
\usepackage{bm} 
\usepackage{amsfonts,amsmath,amssymb,amsbsy,latexsym}

\begin{document}

\markboth{Stability and Diversity in Collective Adaptation}{Sato, Akiyama, and Crutchfield}


\title{Stability and Diversity in Collective Adaptation}

\author{Yuzuru Sato}
\email[Electronic address: ]{ysato@santafe.edu}
\affiliation{Santa Fe Institute, 1399 Hyde Park Road, Santa Fe, NM
87501, USA}

\author{Eizo Akiyama}
\affiliation{Institute of Policy and Planning Sciences, University of Tsukuba,\\
Tennodai 1-1-1, Tsukuba, Ibaraki 305-8573, Japan}

\author{James P. Crutchfield}
\affiliation{Santa Fe Institute, 1399 Hyde Park Road, Santa Fe, NM
87501, USA}


\begin{abstract}
We derive a class of macroscopic differential equations that describe
collective adaptation, starting from a discrete-time stochastic microscopic
model. The behavior of each agent is a dynamic balance between adaptation
that locally achieves the best action and memory loss that leads to
randomized behavior. We show that, although individual agents interact
with their environment and other agents in a purely self-interested way, 
macroscopic behavior can be interpreted as game dynamics. Application
to several familiar, explicit game interactions shows that the adaptation
dynamics exhibits a diversity of collective behaviors. 
The simplicity of the assumptions underlying the macroscopic equations
suggests that these behaviors should be expected broadly in collective
adaptation. We also analyze the adaptation dynamics from an
information-theoretic viewpoint and discuss self-organization 
induced by information flux between agents, giving a novel view of collective
adaptation.
\end{abstract}



\maketitle



\section{Introduction}

Collective behavior in groups of adaptive systems is an important and
cross-cutting topic that appears under various guises in many fields,
including biology, neurosciences, 
computer science, and social science. 
In all these adaptive systems, individual agents interact with 
one another and modify their behaviors according to the information they
receive through those interactions. Often, though, collective behaviors emerge 
that are beyond the individual agent's perceptual capabilities and that 
sometimes frustrate satisfying the local goals. With competitive interactions
dynamic adaptation can produce rich and unexpected behaviors. This kind of
mutual adaptation has been discussed, for example, in studies of biological
group interaction \cite{Win80,Hof88,Cama01a}, interactive learning
\cite{Bat87,Ros87,Tai99}, large-scale adaptive systems
\cite{Simo96a,Broo95a,Holl92a}, and learning in games \cite{Bor97,Fud98}.

Here we develop a class of coupled differential equations for mutual
adaptation in agent collectives---systems in which agents learn how to
act in their environment and with other agents through reinforcement of
their actions. We show that the adaptive behavior in agent collectives,
in special cases, reduces to a generalized form of multipopulation replicator
equations and, generally, can be viewed as a kind of information-theoretic
self-organization in a collective adaptive system.

Suppose that many agents interact with an environment and each independently
attempts to adjust its behavior to the environment based on its sensory
stimuli. The environment consists of other agents and other exogenous
influences. The agents could be humans, animals, or machines, but we make no
assumptions about their detailed internal structures. That is, the central
hypothesis in the following is that collective adaptation is a dynamical
behavior driven by agents' environment-mediated interactions. 
By separating the time scales of change in the environment, of 
agents' adaptation, and of agent-agent interactions, our models describe,
not the deterministic decision-making itself, but the temporal change in the
probability distribution of choices.

\subsection{Related Work}

This approach should be compared and contrasted with game theoretic
view \cite{Neu44}. First, classical game theory often assumes that 
players have knowledge of the entire environmental structure 
and of other players' decision-making processes. 
Our adaptive agents, however, have no knowledge of a game in which
they might be playing. Thus, unlike classical game theory, in our setting there
is no bird's eye view for the entire collective that is available to
the agents. Agents have only a myopic model of the environment,  
since any information external to them is given 
implicitly via the reinforcements for their
action choices. Second, although we employ 
game-theoretic concepts such as Nash equilibria, we focus almost
exclusively on \emph{dynamics}---transients, attractors, and so
on---of collective adaptation, while, naturally, making contact with
the \emph{statics} familiar from game theory. Finally, despite the
differences, game structures can be introduced as a set of parameters 
corresponding to approximated static environments. 

While replicator dynamics were introduced originally for evolutionary game
theory \cite{Tay78,Tay79,Weib95a}, the relationship between learning
with reinforcement and replicator equations has been discussed only 
recently \cite{Bor97,Fud98}. Briefly stated, in our model the state space
represents an individual agent's probability distribution to choose 
actions and the adaptation equations describe the temporal evolution of
choice probabilities as the agents interact. Here, we extend these
considerations to collective adaptation, introducing the theory behind a
previously reported model \cite{Sat02,Sat03}. The overall approach, though,
establishes a general framework for dynamical-systems modeling and analysis
of adaptive behavior in collectives. It is important to emphasize that
our framework goes beyond the multipopulation replicator equations and
asymmetric game dynamics since it does not require a static environment
(cf. Ref. \cite{Akiy00a,Akiy02a} for dynamic environments) and it includes the
key element of the temporal loss of memory.

We model adaptation in terms of the distribution of agents' choices,
developing a set of differential equations that are a continuous-time
limit of a discrete-time stochastic process; cf. Ref. \cite{Ito79}.
We spend some time discussing the origin of action probabilities, since
this is necessary to understand the model variables and also to clarify
the limits that we invoke to arrive at our model. One is tempted to give a
game-theoretic interpretation of the model and its development. For example,
the mixed strategies in game play are often interpreted as weights over all
(complete plans of) actions. However, the game-theoretic view 
is inappropriate for analyzing local, 
myopic adaptation and the time evolution of collective behavior. 

Another interpretation of our use of action probabilities comes from regarding
them as frequencies of action choices. In this view, one needs long-time
trials so that the frequencies take on statistical validity for an agent.
Short of this, they would be dominated by fluctuations, due to undersampling.
In particular, one requires that stable limit distributions exist. Moreover,
the underlying deterministic dynamics of adaptation should be ergodic and have strong mixing
properties. Finally, considering agent-agent interactions, one needs
to assume that their adaptation is very slow compared to interaction dynamics. 
For rapid, say, real-time adaptation, these assumptions would be invalid.
Nonetheless, they are appropriate for long-term reinforcement, as found in
learning motion through iterated exercise and learning customs through 
social interaction. 

\subsection{Synopsis}

The approach we take is ultimately phenomenological. We are reminded of
the reaction-diffusion models of biological morphogenesis introduced
originally in Ref. \cite{Tur52}. There, the detailed processes of biological
development and pattern formation were abstracted, since their biochemical
basis was (and still is) largely unknown, and a behavioral phenomenology was
developed on this basis. Similarly, we abstract the detailed and unknown
perceptual processes that underlie agent adaptation and
construct a phenomenology that captures adaptive behavior at a
larger scale, in agent collectives. 

The phenomenology that we develop for this is one based on communications
systems. Agents in a collective are confronted with the same three
problems of communication posed by Weaver in the founding work of information
theory---\emph{The Mathematical Theory of Communication} \cite{Sha49}: 
(a)``How accurately can the symbols of communication be
transmitted?'', (b)``How precisely do the transmitted symbols convey the desired
meaning?'' and (c)``How effectively does the received meaning affect conduct
in the desired way?''.  
Shannon solved 
the first problem developing his theory of error-free transmission
\cite{Sha49}. In their vocabulary adaptive agents are \emph{information sources}. 
Each (a) receives information transmitted from the external
environment, which includes other agents, 
(b) interprets the received information and modifies its
internal model accordingly, and then, (c) making decisions based on
the internal model, generates future behavior.

We will show that this information-theoretic view provides useful tools
for analyzing collective adaptation and also an appropriate description
for our assumed frequency dynamics. Using these we derive a new state
space based on the self-informations of agent's actions and this allows
one to investigate the dynamics of uncertainty in collective adaptation. 
It will become clear, though, that the assumption of global information 
maximization has 
limited relevance here, even for simple mutual adaptation in a static
environment. Instead, self-organization that derive from 
the information flux 
between agents gives us a new view of collective adaptation. 

To illustrate collective adaptation, we present several simulations of
example environments; in particular, those having frustrated agent-agent
interactions \cite{McC45}. Interestingly, for two agents with perfect
memory interacting via zero-sum rock-scissors-paper interactions the
dynamics exhibits Hamiltonian chaos \cite{Sat02}. With memory loss,
though, the dynamics becomes dissipative and displays the full range
of nonlinear dynamical behaviors, including limit cycles, intermittency,
and deterministic chaos \cite{Sat03}. 

The examples illustrate that Nash equilibria 
often plays little or no role in collective adaptation. 
They are fixed points determined by the intersections of nullclines of the 
adaptation dynamics and sometimes 
the dynamics is explicitly excluded from reaching 
Nash equilibria, even asymptotically. 
Rather, it turns out that the network describing the switching 
between deterministic actions is a dominant factor in structuring the
state-space flows. From it, much of the dynamics, including the origins
of chaos becomes intuitively clear.

In the next section (Sec. \ref{Sec:Dynamics}), we develop a dynamical system
that models adaptive behavior in collectives. In Sec. \ref{Sec:InfoSpace} we
introduce an information-theoretic view and coordinate-transformation for
adaptation dynamics and discuss self-organization 
induced by information flux. To illustrate the rich range of
behaviors, in the Sec. \ref{Sec:Examples} 
we give several examples of adaptive dynamics based on 
non-transitive interactions. Finally, in Sec. \ref{Sec:TheEnd} 
we interpret our results and suggest future directions.

\section{Dynamics for Collective Adaptation}
\label{Sec:Dynamics}

Before developing the full equations for a collective of adaptive agents,
it is helpful to first describe the dynamics of how an individual agent
adapts to the constraints imposed by its environment using the memory of its
past behaviors. We then build up a description of how multiple agents interact,
focusing only on the additional features that come from interaction. The result
is a set of coupled differential equations that determine the behavior of
adaptive agent collectives and are amenable to various kinds geometric,
statistical, and information-theoretic analyses.

\subsection{Individual Agent Adaptation} 

Here we develop a continuous-time model for adaptation in an environment
with a single adaptive agent. Although the behavior in this case is relatively
simple, the single-agent case allows us to explain several basic points about 
dynamic adaptation, without the complications of a collective and agent-agent
interactions. In particular, we discuss how and why we go from a discrete-time
stochastic process to a \emph{continuous-time} limit. We also describe an
agent's effective internal model of the environment and how we model its
adaptation process via a \emph{probability distribution} of action choices.

An agent takes one of $N$ possible \emph{actions}: $i = 1, 2, \ldots, N$
at each time step $\tau$. Let the probability for the agent to chose action
$i$ be $x_i(\tau)$, where $\tau$ is the number of steps from the initial
state $x_i(0)$. The agent's state vector---its \emph{choice
distribution}---at time $\tau$ is ${\bf x}(\tau)=(x_1(\tau),x_2(\tau),\ldots,x_N(\tau))$, where 
$\Sigma_{n=1}^N x_n(\tau) = 1$. In the following we call the temporal behavior of
${\bf x} (\tau)$ as the \emph{dynamics of adaptation}.

Let $r_i(\tau)$ denote the reinforcement the agent receives for its
taking action $i$ at step $\tau$. Denote the collection of these 
by the vector ${\bf r}(\tau) = (r_1(\tau), \ldots, r_N(\tau))$. 
The agent's memories---denoted ${\bf
Q}(\tau)=(Q_1(\tau),\ldots,Q_N(\tau))$
---of past rewards from its actions are updated according to 
\begin{equation}
Q_i(\tau+1)-Q_i(\tau)  = 
	\frac 1T\left[\delta_{i}(\tau) r_{i}(\tau) - \alpha Q_i
	(\tau)\right] ~,
\label{SingleMemoryUpdate}
\end{equation}
where
\begin{equation}
\delta_{i}(\tau)  = 
	\left\{
	\begin{array}{l}
		1, ~\mbox{action $i$ chosen at step $\tau$}\\
		0, ~\mbox{otherwise}\\
	\end{array}
	\right. 
\end{equation}
with $i = 1, \ldots, N$ and $Q_i(0)= 0$. $T$ is a constant that sets the
agent-environment interaction time scale. $\alpha \in [0,1)$ controls the 
agent's memory loss rate. For $\alpha=0$, the agent has a perfect memory
as the sum of the past reinforcements; for $\alpha>0$ the memory is attenuated
in that older reinforcements have less effect on the current $Q_i$s and more
recent reinforcements are given larger weight. One imagines that the agent
constructs a histogram of past reinforcements and this serves as a simple
internal memory of its environment. 

An agent chooses its next action according to its choice distribution which
is updated from the reinforcement memory according to: 
\begin{equation}
x_i (\tau) = \frac{e^{\beta Q_i (\tau)}} {\sum_{n=1}^{N}
  e^{\beta Q_n (\tau)}} ~,
\label{SingleVector}
\end{equation}
where $i=1, 2, \ldots, N$. 
$\beta\in [0,\infty]$ controls the adaptation rate: how much the
choice distribution is changed by the memory of past reinforcements. 
For example, if $\beta = 0$, the choice distribution is unaffected by
past reinforcements. Specifically, it becomes independent of $\bf Q$ and
one has $x_i (\tau) = 1/N$. In this case, the agent chooses actions with
uniform probability and so behaves completely randomly. 
In a complementary fashion, in the limit $\beta \rightarrow \infty$, an agent chooses
that action $i$ with the maximum $Q_i (\tau)$ and 
$x_i(\tau) \rightarrow 1$. 

Given Eq. (\ref{SingleVector}) the time evolution of agent's choice
distribution is:  
\begin{equation}
x_i (\tau+1) = \frac{x_i (\tau) e^{\beta (Q_i(\tau+1)-Q_i(\tau))}}
  {\sum_{n=1}^{N} x_n (\tau) e^{\beta (Q_n(\tau+1)-Q_n (\tau))}} ~, 
\label{SingleVectorUpdate} 
\end{equation}
where $i = 1, 2, \ldots, N$. 
This determines how the agent adapts its choice distribution using
reinforcements it has received from the environment for its past actions. 

This simple kind of adaptation was introduced as a principle of behavioral
learning \cite{Skin38,Hebb49a} and as a model of stochastic learning 
\cite{Nor72}, and is sometimes referred to as reinforcement learning
\cite{Sam67,Sutt98a}. Arguably, it is the simplest form of adaptation in which
an agent develops relationships or behavior patterns through reinforcements
from external stimuli. 

\begin{figure}[htbp]
 \begin{center}
  \leavevmode
  \includegraphics[scale=0.6]{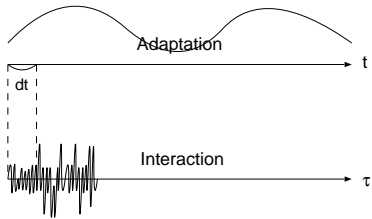}
 \end{center}
 \caption{The time scale ($t$) of a single agent interacting with its
	environment and the time scale ($\tau$) of the agent's 
	adaptation: $\tau \ll t$.
	}
 \label{fig:Timescales}
\end{figure}

Starting with the discrete-time model above, one can develop a continuous-time
model that corresponds to the agent performing a large number of actions,
iterates of Eq. (\ref{SingleMemoryUpdate}), 
for each choice distribution
update, iterate of Eq. (\ref{SingleVector}). Thus, we recognize two
different time scales: one for agent-environment interactions and one for
adaptation of the agent's internal model based on its internal memory. 
We assume that the adaptation dynamics is very slow compared to interactions
and so $\bf x$ is essentially constant during interactions. (See Fig.
\ref{fig:Timescales}.)   

Starting from Eq. (\ref{SingleMemoryUpdate}), 
one can show that the 
continuous-time dynamics of memory updates is given by the differential
equations 
\begin{equation}
\dot{Q}_i(t) = R_i(t) - \alpha Q_i(t) ~, 
\label{SingleMemoryUpdate-Continuous}
\end{equation} 
with $i = 1, 2, \ldots, N$ and $Q_i (0) = 0$. (see App.
\ref{ContinuousTimeLimits}.) Here $R_i$ is the reward the environment gives
to the agent choosing action $i$: the average of $r_i (\tau)$ during the time
interval between updates of $\mathbf{x}$ at $t$ and $t + dt$.

From Eq. (\ref{SingleVector}) one sees that the map from ${\bf Q}(t)$ to
$\mathbf{x}(t)$ at time $t$ is given by
\begin{equation}
x_i (t) = \frac{e^{\beta Q_i (t)}} {\sum_{n=1}^N e^{\beta Q_n (t)}} ~, 
\label{SingleVector-Continuous} 
\end{equation}
where $i = 1, 2, \ldots, N$. 
Differentiating Eq. (\ref{SingleVector-Continuous}) gives
the continuous-time dynamics
\begin{equation} 
\dot{x}_i(t) = \beta x_i(t) (\dot{Q}_i(t)
  - \sum_{n=1}^N \dot{Q}_n(t) x_n(t)) ~,
\label{SingleVectorUpdate-Continuous}
\end{equation}
with $i = 1, 2, \ldots, N$. 

Assembling Eqs. (\ref{SingleMemoryUpdate-Continuous}), 
(\ref{SingleVector-Continuous}), and (\ref{SingleVectorUpdate-Continuous}),
one finds the basic dynamic that governs agent behavior on the adaptation
time-scale:
\begin{equation} 
\frac{\dot{x_i}}{x_i} = \beta ( R_i - R ) + \alpha ( H_i - H ) ~,
\label{SingleLearningDynamics-Continuous}
\end{equation}
where $i = 1, 2, \ldots, N$. Here
\begin{equation}
R=\sum_{n=1}^N x_n R_n
\end{equation}
is the net reinforcement averaged over the agent's possible actions.
And, 
\begin{equation}
H_i = -\log x_i 
\end{equation}
where $i = 1, 2, \ldots, N$, is the \emph{self-information} or degree of surprise
when the agent takes action $i$ \cite{Sha49}. The average self-information,
or \emph{Shannon entropy} of the choice distribution, also appears as
\begin{equation}
H = \sum_{n=1}^N x_n H_n = -\sum_{n=1}^N x_n \log x_n ~.
\end{equation}
These are the entropies of the agent's choice distribution measured, 
not in \emph{bits} (binary digits), but in \emph{nats} (natural digits),  
since the natural logarithm is used. The entropy measures the choice
distribution's flatness, being maximized when the choices all have equal
probability. 

Fortunately, the basic dynamic captured by Eq.
(\ref{SingleLearningDynamics-Continuous}) is quite intuitive, being the
balance of two terms on the right-hand side. The first term describes an
adaptation dynamic, whose time scale is controlled by $\beta$. The second
describes the loss of memory with a time scale controlled by $\alpha$.
That is, the adaptation in choice probabilities is driven by a balance
between two forces: {the tendency to concentrate the choice probability 
based on the reinforcement ${\bf R}=(R_1, R_2, \ldots, R_N)$ and the 
tendency to make choices equally likely.} 
Finally, on the lefthand side, one has the logarithmic
derivative of the choice probabilities: $\dot{x}_i/x_i = d/dt ~(\log x_i)$.

Note that each of the terms on the righthand side is a difference between
a function of a particular choice and that function's average. Specifically,
the first term $\Delta R_i \equiv R_i - R$ is the relative benefit in choosing
action $i$ compared to the mean reinforcement across all choices. Other 
things being held constant, if this term is positive, then action $i$ is the
better choice compared to the mean and $x_i$ will increase. The second term
$\Delta H_i\equiv H_i - H$ is the relative informativeness of 
taking action $i$ compared to the average $H$, that is Shannon
entropy. Thus, $x_i$ decreases in proportion to the entropy at time 
$t$ and so this term works to increase the uncertainty of agent's
actions, flattening the choice 
distribution by increasing the probability of unlikely actions. When
$x_i = N^{-1}$, the distribution is flat (purely random choices),
$\Delta H = 0$, and memory loss effects disappear. 

Mathematically, the adaptation equations have quite a bit of structure
and this has important consequences, as we will see.
Summarizing, the adaptation equations describe a dynamic that balances 
the tendency to concentrate on choices associated with the 
best action against the tendency to make the
choices equally likely. The net result is to increase 
the choice uncertainty, subject to the constraints 
imposed by the environment via the 
reinforcements. Thus, the choice distribution is 
the least biased distribution 
consistent with environmental constraints and individual memory loss. 
We will return to discuss this
mechanism in detail using information theory in the Sec. \ref{Sec:InfoSpace}.

\begin{figure}[htbp]
 \begin{center}
  \leavevmode
  \includegraphics[scale=0.45]{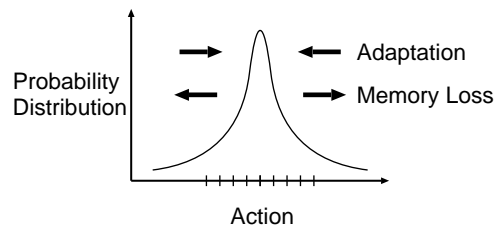}
 \end{center}
\caption{A dynamic balance of adaptation and memory
  loss:  Adaptation concentrates the probability distribution on
  the best action. Memory loss of past history leads to a
  distribution that is flatter and has higher entropy. 
  }
\label{fig:BalanceAdapMeM}
\end{figure}

\begin{figure}[htbp]
 \begin{center}
  \leavevmode
  \includegraphics[scale=0.6]{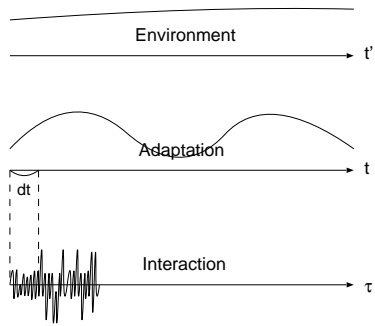}
 \end{center}
\caption{The time scales of dynamic adaptation: Agent adaptation is slow
  compared to agent-environment interaction and environmental change is
  slower still compared to adaptation.
  }
\label{fig:ThreeTimescales}
\end{figure}

Since the reinforcement determines the agent's interactions
with the environment, there are, in fact, three different time scales
operating: that for agent-environment interactions, that for each agent's
adaptation, and that for changes to the environment. However, if the
environment changes very slow compared to the agent's internal adaptation, 
the environment $r_{i}(t)$ can be regarded as effectively constant, as shown
in Fig. \ref{fig:ThreeTimescales}.

In this case $r_i(t)$ can be approximated as a static relationship between 
an agent's actions and the reinforcements given by the environment. Let
$r_i(t) = a_i$, where $\mathbf{a} = ( a_1, \ldots, a_N )$ are constants
that are normalized: $\Sigma_{n=1}^N a_n =0$. Given this, the agent's
time-average reinforcements are $a_i$ ($R_i = a_i$) and the continuous-time 
dynamic simplifies to: 
\begin{equation}
\frac{\dot{x}_i}{x_i} = \beta (a_i-\sum_{n=1}^N a_n x_n)
  + \alpha (-\log x_i + \sum_{n=1}^N x_n\log x_n) ~, 
\label{SingleLearningDynamics-Continuous-Constant}
\end{equation}
where $i = 1, 2, \ldots, N$. 

The behavior of single-agent adaptation given by Eq. 
(\ref{SingleLearningDynamics-Continuous-Constant}) is very simple. When
$\alpha$ is small, so that adaptation is dominant $x_i \rightarrow 1$, where
$i$ is the action with the highest reward $a_i$, and $x_j\rightarrow 0$
for $j \neq i$. The agent receives this information from the fixed environment
and its behavior is simply to choose the action
with the maximum reward and the choice distribution moves to the associated
simplex vertex ${\bf x}^*=(0, \ldots, 1^{\stackrel{i}{\vee}}, \ldots,0)$. 
In the special case when
$\alpha=0$, it is known that for arbitrary $\bf a$ Eq.
(\ref{SingleLearningDynamics-Continuous-Constant}) moves $\mathbf{x}$
to the vertex corresponding to the maximum $a_i$ \cite{Hof88}. In a
complementary way, when $\alpha$ is large enough to overcome the relative
differences in reinforcements---that is, when
$\beta/\alpha\rightarrow 0$ memory loss dominates, the agent
states goes to a uniform choice distribution ($x_i = N^{-1}$) and the
system converges to the simplex center. Note that in machine learning
this balance between local optimization and randomized behavior, which
selects non-optimal actions, is referred to as the
\emph{exploitation-exploration trade-off} \cite{Sutt98a}.

\begin{figure}[htbp]
  \begin{center}
    \leavevmode
	\includegraphics[scale=0.55]{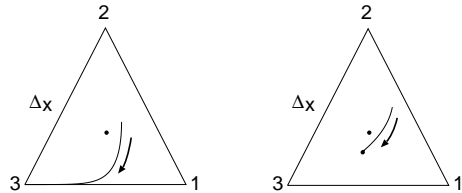}
  \end{center}
\caption{Dynamics of single-agent adaptation: Here there are three actions,
  labeled $1$, $2$, and $3$, and the environment gives reinforcements according
  to ${\bf a}=(\frac23\epsilon, -1-\frac13\epsilon, 1-\frac13\epsilon)$.
  The figure shows two trajectories from simulations with $\epsilon = 0.5$
  and $\beta=0.1$ and with $\alpha = 0.0$ (right) and $\alpha = 0.3$ (left). 
}
\label{fig:SingleLearningTrajectories}
\end{figure}

For instance, consider an agent that takes $N = 3$ actions, $\{1,2,3\}$, in an
environment described by
${\bf a}=(\frac23\epsilon, -1-\frac13\epsilon,1-\frac13\epsilon)$,
with $\epsilon \in [-1, 1]$. In the perfect memory case ($\alpha=0$), 
the choice distribution converges to 
a stable fixed point $(0,0,1)$.  ${\bf x}^*=(\frac13, \frac13, \frac13)$ 
is an unstable hyperbolic fixed point. In the memory loss case
($\alpha>0$), dynamics converges a stable fixed point inside the simplex. 
(These cases are illustrated in Fig. \ref{fig:SingleLearningTrajectories}.)

Even when the environment is time-dependent, the agent's behavior can track
the highest-reward action as long as the time scale of environment change
is slow compared to the agent's adaptation. However, the situation is more
interesting when environment change occurs at a rate near the time-scale set 
by adaptation. Mutual adaptation in agent collectives, the subject of the
following sections, corresponds to just this situation. Other agents
provide, thought their own adaptation, a dynamic environment to any given
agent and if their times scales of adaptation are close the dynamics
can be quite rich and difficult to predict and analyze.

\subsection{Two Agent Adaptation}

To develop equations of motion for adaptation in an agent collective we
initially assume, for simplicity, that there are only two agents. The
agents, denoted $X$ and $Y$, at each moment take one of $N$ or $M$ actions,
respectively. The agents states at time $t$ are ${\bf x}=(x_1, \ldots,x_N)$
and ${\bf y} = (y_1, \ldots, y_M)$, with
$\Sigma_{n=1}^N x_n = \Sigma_{m=1}^M y_m = 1$. ${\bf x}(0)$ and ${\bf y}(0)$
are the initial conditions. We view the time evolution of each agent's state
vector in the simplices ${\bf x}\in \Delta_X$ and ${\bf y} \in \Delta_Y$ and
the group dynamics in the \emph{collective} state space $\Delta$ which is the
product of the agent simplices: 
\begin{equation}
{\bf X}=({\bf x},{\bf y}) \in \Delta = \Delta_X \times \Delta_Y ~.
\end{equation}
There are again three different time scales to consider: one for agent-agent
interaction, one for each agent's internal adaptation, and one for the
environment which now mediates agent interactions via the reinforcements
given to the agents. Here we distinguish between the \emph{global environment}
experienced by the agents and the \emph{external environment}, which is the
global environment with the agent states removed. The external environment
controls, for example, the degree of coupling between the agents. In contrast
with the single-agent case, in the many agent setting each agent's behavior
produces a dynamic global environment for the other. This environment dynamics is
particularly important when the adaptation time scales of each agent are close.

Following the single-agent case, though, we assume that the adaptation
dynamic is very slow compared to that of agent-agent interactions and that
the dynamics of the external environment changes very slowly compared to
that of agents' mutual adaptation. Under these assumptions the agent state
vectors $\mathbf{x}$ and $\mathbf{y}$ are effectively constant during
the agent-agent interactions that occur between adaptation updates. The
immediate consequence is that can describes the collective state space in
terms of the frequencies of actions (the choice distributions). Additionally,
the environment is essentially constant relative to changes in the states
$\mathbf{x}$ and $\mathbf{y}$.

Denote the agents' memories by ${\bf Q}^X=(Q_1^X,\ldots, Q_N^X)$ for $X$
and ${\bf Q}^Y=(Q_1^Y,\ldots,Q_M^Y)$ for $Y$ and set $Q_i^X(0)=0$ and
$Q_j^Y(0)=0$, for for $i=1,\ldots, N$ and $j=1,\ldots,M$. For the dynamic
governing memory updates we have 
\begin{eqnarray}
Q_i^X(\tau+1)-Q_i^X(\tau)  &=&
	\frac 1T\left[\delta_{ij}(\tau) r^X_{ij}(\tau) - \alpha_X Q_i^X
	(\tau)\right] ~, \nonumber\\
Q_j^Y(\tau+1)-Q_j^Y(\tau)  &=&
	\frac 1T\left[\delta_{ij}(\tau) r^Y_{ji}(\tau) - \alpha_Y Q_j^Y
	(\tau)\right] ~, \nonumber\\
\label{MultiMemoryUpdate}
\end{eqnarray}
where
\begin{equation}
\delta_{ij}(\tau)  = 
	\left\{
	\begin{array}{l}
		1, ~\mbox{pair of actions $(i, j)$ chosen at step $\tau$}\\
		0, ~\mbox{otherwise}\\
	\end{array}
	\right. 
\end{equation}
with $i = 1, \ldots, N$, $j = 1, \ldots, M$  
and $Q_i^X(0)= 0$, $Q_j^Y(0)= 0$. $T$ is a time constant. 
Then the continuous-time dynamics of memory updates for $X$ and $Y$ 
are given by the differential equations 
\begin{eqnarray} 
\dot{Q}_i^X &=& R_i^X - \alpha_X Q_i^X ~,\nonumber\\
\dot{Q}_j^Y &=& R_j^Y - \alpha_Y Q_j^Y ~, 
\label{MultiMemoryUpdate-Continuous}
\end{eqnarray}
for $i=1, 2, \ldots, N$ and $j=1, 2, \ldots, M$. 
$R_i^X$ is the reward for agent $X$ choosing action $i$, averaged over agent
$Y$'s actions between adaptive updates; and $R_j^Y$ is $Y$'s. The parameters
$\alpha_X, \alpha_Y \in [0,1)$ control each agent's memory loss rate,
respectively.

The map from $\mathbf{Q}^X (t)$ to $\mathbf{x} (t)$ and 
from ${\bf Q}^Y (t)$ to $\mathbf{y} (t)$ at time $t$ is
\begin{eqnarray}
x_i (t) &=& \frac{e^{\beta_X Q_i^X (t)}}
	{\sum_{n=1}^N e^{\beta_X Q_n^X(t)}} ~,\nonumber\\
y_j (t) &=& \frac{e^{\beta_Y Q_j^Y (t)}}
	{\sum_{m=1}^M e^{\beta_Y Q_m^Y(t)}} ~, 
\label{Vector-Continuous}
\end{eqnarray}
for $i = 1, \ldots, N$ and $j = 1, \ldots, M$. Here
$\beta_X, \beta_Y\in [0,\infty]$ control the agents' adaptation rates,
respectively. Differentiating Eq. (\ref{Vector-Continuous}) with
respect to $t$, the continuous-time adaptation for two agents is governed by
\begin{eqnarray}
\dot{x}_i &=&
	\beta_X x_i (\dot{Q}_i^X - \sum_{n=1}^N \dot{Q}^X_n x_n) ~,\nonumber\\
\dot{y}_j &=&
	\beta_Y y_j (\dot{Q}_j^Y - \sum_{m=1}^M \dot{Q}^Y_m y_m) ~,
\label{VectorUpdate-Continuous}
\end{eqnarray}
for $i = 1, \ldots, N$ and $j = 1, \ldots, M$.

Putting together Eqs. (\ref{MultiMemoryUpdate-Continuous}), 
(\ref{Vector-Continuous}), and (\ref{VectorUpdate-Continuous}),
one finds the coupled adaptation equations for two agents:
\begin{eqnarray} 
\frac{\dot{x_i}}{x_i}  
	& = & \beta_X (R_i^X - R^X) + \alpha_X (H_i^X - H^X) ~,\nonumber\\
\frac{\dot{y_j}}{y_j}  
	& = & \beta_Y (R_j^Y - R^Y) + \alpha_Y (H_j^Y - H^Y) ~, \nonumber\\
\label{LearningEquations}
\end{eqnarray}
for $i = 1, \ldots, N$ and $j = 1, \ldots, M$ and where 
\begin{eqnarray}
R^X= \sum_{n=1}^N x_n R_n^X,&& R^Y= \sum_{m=1}^M y_m R_m^Y ~, \nonumber\\
H^X=\sum_{n=1}^N x_n H_n^X,&& H^Y=\sum_{m=1}^M y_m H_m^Y ~. 
\end{eqnarray}
The interpretations of the $\Delta R = R_i - R$ and $\Delta H = H_i - H$
terms are not essentially different from those introduced to describe the
single-agent case. That is, the behavior of each agent is a dynamic balance
between (i) adaptation: concentrating the choice probability on the best action
at $t$ and (ii) memory loss: increasing the choice uncertainty. What is new here is that
there are two (and eventually more) agents attempting to achieve this balance
together using information that comes from their interactions with the
global environment.

As given, the adaptation equations include the possibility of a time-dependent
environment, which would be implemented, say, using a time-dependent
reinforcement scheme. However, as with the single-agent case, it is helpful to
simplify the model by assuming a static external environment and, in particular, static
relationships between the agents.

Assume that the external environment changes slowly compared to the dynamics
of mutual adaptation, as illustrated in Fig. \ref{fig:ThreeTimescales}. This
implies a nearly static relationship between pairs of action choices $(i,j)$ 
and reinforcements $r^X_{ij}$ and $r^Y_{ji}$ 
for both agents. Since the environmental dynamics
is very slow compared to each agents' adaptation, $r_{ij}^X(t)$ and
$r_{ji}^Y(t)$ are essentially constant during adaptation. The $r$s can be
approximated then as constant:
\begin{eqnarray}
r_{ij}^X(t) &=& a_{ij} ~,\nonumber \\
r_{ji}^Y(t) &=& b_{ji} ~,  
\label{NormalRewards}
\end{eqnarray}
for $i = 1, \ldots, N$ and $j = 1, \ldots, M$. 
$a_{ij}$ and $b_{ji}$ are normalized over $j$ and $i$ so that when summing
over all actions the reinforcements vanish:
\begin{eqnarray}
\sum_{n=1}^N a_{nj}=0 ~,\nonumber \\
\sum_{m=1}^M b_{mi}=0 ~.  
\end{eqnarray}
Given the form of $\Delta R$ in the adaptation equations, this normalization
does not affect the dynamics. 

Assume further that $\bf x$ and $\bf y$ are independently distributed.
This is equivalent to agents never having a global view of the
collective or their interactions with 
the environment (other agents). Each agent's knowledge of the environment
is uncorrelated, at each moment, with the state of the other agents.
The time-average rewards for $X$ and $Y$ now become 
\begin{eqnarray}
  R_i^X &=& \sum_{m=1}^M a_{im} y_m=(A{\bf y})_i ~,\nonumber\\
 R_j^Y &=&\sum_{n=1}^N b_{jn} x_n=(B{\bf x})_j ~, 
\label{ConstantRewards}
\end{eqnarray}
for $i = 1, \ldots, N$ and $j = 1, \ldots, M$.
In this restricted case, the continuous-time dynamic is given by the coupled
adaptation equations
\begin{eqnarray} 
\frac{\dot{x}_i}{x_i} & = &
	\beta_X [ (A{\bf y})_i-{\bf x}\cdot A{\bf y}] \nonumber\\
	&+& \alpha_X [-\log x_i+\sum_{n=1}^N x_n \log x_n] ~, \nonumber\\
\frac{\dot{y}_j}{y_j} & = &
	\beta_Y [ (B{\bf x})_j-{\bf y}\cdot B{\bf x}] \nonumber\\
	&+& \alpha_Y [-\log y_j+\sum_{m=1}^M y_m \log y_m] ~. 
\label{LearningEquations-Constant}
\end{eqnarray}
for $i = 1, \ldots, N$ and $j = 1, \ldots, M$. $A$ is an $N \times M$ matrix
and $B$ is a $M \times N$ matrix with $(A)_{ij} = a_{ij}$ and
$(B)_{ji} = b_{ji}$, respectively. ${\bf x}\cdot A{\bf y}$ is the inner
product between ${\bf x}$ and $A{\bf y}$ and similarly for 
${\bf y}\cdot B{\bf x}$:
\begin{eqnarray}
{\bf x}\cdot A{\bf y}&=&\sum_{n=1}^N\sum_{m=1}^M a_{nm} x_n y_m
 ~,\nonumber\\
{\bf y}\cdot B{\bf x}&=&\sum_{m=1}^M\sum_{n=1}^N b_{mn} y_m x_n ~. 
\end{eqnarray}

\subsection{Collective Adaptation}

Generalizing to an arbitrary number of agents at this point should
appear straightforward. It simply requires extending Eqs.
(\ref{LearningEquations}) to a collection of adaptive agents. Suppose
there are $S$ agents labeled $s=1,2,\ldots,S$ and each agent can
take one of $N^s$ actions. 
One describes the time evolution of the agents' state vectors in the
simplices ${\bf x}^1 \in \Delta_1$, ${\bf x}^2 \in \Delta_2$, ...,
and ${\bf x}^S\in \Delta_S$. The adaptation dynamics in the
higher-dimensional {\em collective} state space occurs within
\begin{equation}
{\bf X}=({\bf x}^1,{\bf x}^2, \ldots, {\bf x}^S) \in
	\Delta = \Delta_1 \times \Delta_2 \times \ldots \Delta_S ~.
\end{equation}
Then we have the dynamics for collective adaptation as 
\begin{equation} 
\frac{\dot{x_{i^s}^s}}{x_{i^s}^s}  
	= \beta_s ( R_{i^s}^s - R^s) +\alpha_s (H_{i^s}^s - H^s) ~.
\label{MultiLearningEquations}
\end{equation}
for $i^s = 1, \ldots, N^s$ and $s = 1, \ldots, S$. 
$R_{i^s}^s$ and $H_{i^s}^s$ 
are the reinforcement and the self-information for $s$ to choose
action $i^s$, respectively. 
Equations (\ref{MultiLearningEquations}) constitute our general model for
adaptation in agents collective.

With three agents $X$, $Y$, and $Z$, with collective state space
\begin{equation}
{\bf X}=({\bf x}, {\bf y}, {\bf z}) \in
	\Delta = \Delta_X \times \Delta_Y \times \Delta_Z ~.
\end{equation}
one obtains: 
\begin{eqnarray} 
\frac{\dot{x_i}}{x_i}  
	& = & \beta_X (R_i^X - R^X) + \alpha_X [H_i^X-H^X] ~, \nonumber\\
\frac{\dot{y_j}}{y_j}  
	& = & \beta_Y (R_j^Y - R^Y) 	+ \alpha_Y [H_j^Y-H^Y] ~, \nonumber\\
\frac{\dot{z_k}}{z_k}  
	& = & \beta_Z (R_k^Z - R^Z) 	+ \alpha_Z [H_k^Z-H^Z] ~,
\label{MultiLearningDynamics-Continuous-For3}
\end{eqnarray}
for $i = 1, \ldots, N$, $j = 1, \ldots, M$, and $k = 1, \ldots, L$.
The static environment version reduces to 
\begin{eqnarray}
\frac{\dot{x}_i}{x_i} &=& \beta_X [(A{\bf y}{\bf z})_i
	- {\bf x}\cdot A{\bf y}{\bf z}]  \nonumber\\
&+& \alpha_X [-\log x_i+\sum_{n=1}^N x_n \log x_n] ~, \nonumber\\
\frac{\dot{y}_j}{y_j} &=& \beta_Y [(B{\bf z}{\bf x})_j
	- {\bf y}\cdot B{\bf z}{\bf x}]  \nonumber\\
&+& \alpha_Y [-\log y_j+\sum_{m=1}^M y_m \log y_m] ~, \nonumber\\
\frac{\dot{z}_k}{z_k} &=& \beta_Z [(C{\bf x}{\bf y})_k
	-{\bf z}\cdot C{\bf x}{\bf y}]  \nonumber\\
  &+& \alpha_Z [-\log z_k+\sum_{l=1}^L z_l \log z_l] ~, 
\end{eqnarray}
for $i = 1, \ldots, N$, $j = 1, \ldots, M$, and $k = 1, \ldots, L$,
and with tensors $(A)_{ijk} = a_{ijk}$, $(B)_{jki} = b_{jki}$, 
$(C)_{kij} = c_{kij}$. Here 
\begin{equation}
(A{\bf yz})_i=\sum_{m=1}^M \sum_{l=1}^L a_{iml}y_mz_l 
\end{equation}
and 
\begin{equation}
{\bf x}\cdot A{\bf yz}=\sum_{n=1}^N\sum_{m=1}^M \sum_{l=1}^L a_{nml} x_n y_m z_l
\end{equation}
and similarly for $Y$ and $Z$. 
Note that the general model includes heterogeneous network settings 
with local interactions besides global interactions; see App. \ref{NetworkInteractions}.

\subsection{Evolutionary Dynamics and Game Theory}

We now interrupt the development to discuss the connections between the model
developed this far and models from population dynamics and game theory. There
are interesting connections and also some important distinctions that need
to be kept in mind, before we can move forward.

The special case that allows us to make contact with evolutionary dynamics
and game theory is the restriction to agents with perfect memory interacting
in a static environment. (For further details see App.
\ref{NashEquilibria}.)
In the two agent, static external environment 
case we set $\alpha_X = \alpha_Y = 0$ and equal adaptation
rates, $\beta_X = \beta_Y$. Under these assumptions our model, Eqs.
(\ref{LearningEquations-Constant}), reduces to what is either called
\emph{multipopulation replicator equations} \cite{Tay79} or
\emph{asymmetric game dynamics} \cite{Tay79,Bor97,Fud98}. The equations are:
\begin{eqnarray}
\frac{\dot{x}_i}{x_i} &=&
  (A{\bf y})_i-{\bf x}\cdot A{\bf y} ~,\nonumber\\
\frac{\dot{y}_j}{y_j} &=&
  (B{\bf x})_j-{\bf y}\cdot B{\bf x} ~. 
\label{MultiPopulationReplicator}
\end{eqnarray}
From the perspective of game theory, one regards the interactions determined
by $A$ and $B$, respectively, as $X$'s and $Y$'s \emph{payoff matrices} for a
linear game in which $X$ plays action $i$ against $Y$'s action $j$.
Additionally, ${\bf x}$ and ${\bf y}$, the agent state vectors, are interpreted
as the \emph{mixed strategies}. In fact,
${\bf x} \cdot A{\bf y}$ and ${\bf y}\cdot B{\bf x}$ in Eqs.
(\ref{MultiPopulationReplicator}) formally satisfy von Neumann-Morgenstern utilities
\cite{Neu44}. If they exist in the interior of the collective simplices
$\Delta_X$ and $\Delta_Y$, interior Nash equilibria of the game $(A,
B)$ are the fixed points determined by the intersections of the x- and y-nullclines of Eqs.
(\ref{MultiPopulationReplicator}).

One must be careful, though, in drawing parallels between our general dynamic
setting and classical game theory. In the idealized economic agents, it is often
assumed that  agents have knowledge of the entire game structure
and of other agents' decision-making processes. Its central methodology
derives how these \emph{rational players} should act. 
Our adaptive agents, in contrast, have no knowledge of a game in which
they might be playing, only a myopic model of the environment and, even then,
this is given only implicitly via the reinforcements the agents receive from
the environment. In particular, the agents do not know whether they are playing
a game or not, how many agents there are beyond themselves, or even whether
other agents exist or not. Our model of dynamic adaptation under such
constraints is appropriate nonetheless for many real world adaptive systems,
whether animal, human, or economic agent collectives \cite{Kah00}. The 
bi-matrix game $(A, B)$ appears above as a description of 
the collective's global dynamic only under the assumptions 
that the external environment changes very slowly. 

The connection with evolutionary dynamics is formal and comes from the fact
that Eqs. (\ref{MultiPopulationReplicator}) are the well known replicator
equations of population dynamics \cite{Hof88}. However, the interpretation
of the variables is rather different. Population dynamics views $\bf x$ and
$\bf y$ as two separate, but interacting (infinite size) groups. These two
populations are described as distributions of various organismal
phenotypes. The equations of motion determine the evolution of
these populations over generations and through 
interaction. In our model, in contrast, 
$\bf x$ and $\bf y$ represent the probability to choose actions for 
each agents. The equations of motion describe 
their dynamic adaptation to each other through interaction.

Despite the similarities that one can draw in this special case, it is
important to emphasize that our framework goes beyond the 
multipopulation replicator equations and asymmetric game dynamics. 
First, the reinforcement scheme $\bf R$ need not lead to linear interactions. Second, the model does
not require a static environment described by a constant bi-matrix $(A, B)$.
Finally, the occurrence of the memory loss term is
entirely new and not found in game theory or evolutionary dynamics.

\section{Information, Uncertainty, and Dynamic Adaptation}
\label{Sec:InfoSpace}

We now shift away from a dynamical systems view and, as promised
earlier, begin to think of the agent collective as a communication network.
Although, this initially will appear unrelated, we will show that there is
a close connection between the dynamical and information theoretic 
perspectives---connections that have both mathematical and pragmatic
consequences.

We consider the adaptive agents in the collective to be information sources. 
Each agent receives information from its environment, which includes other
agents. Each agent interprets the received information and modifies its
behavior accordingly, changing from ${\bf x}(t)$ to ${\bf x}(t+dt)$. Each
agent generates a series of messages (actions) based on its updated internal
model and introduces this new behavior back into the environment. This is 
a different interpretation of the interaction 
process in the collective which we motivated up to now only as 
a dynamical process. Now we discuss the adaptive dynamics from 
information theoretic viewpoint. 

\subsection{Dynamics in Information Space}

In this section we introduce a new state space that directly represents the
uncertainties of agent actions. First, as before, for clarity we focus
on the two-agent static-environment case, Eqs.~(\ref{LearningEquations-Constant}).
Since the components of the agents' states are probabilities, the quantities 
\begin{eqnarray}
\xi_i &=& -\log x_i  ~,\nonumber\\
\eta_j &=& -\log y_j ~,
\label{CLR1}
\end{eqnarray}
are the \emph{self-informations} of agents $X$ and
$Y$ choosing actions $i$ and $j$, respectively. When $x_i$ is small, for
example, the self-information $\xi_i$ is large since action $i$ is rarely chosen
by agent $X$. Consider the resulting change in coordinates in
${\bf R}_+^{N}\times {\bf R}_+^{M}$:
\begin{equation} 
{\bf \Xi}=(\mbox{\boldmath $\xi$}, \mbox{\boldmath $\eta$}) 
 = (\xi_1,\ldots,\xi_N) \times (\eta_1,\ldots,\eta_M) ~.
\end{equation}
The normalization conditions---$\Sigma_{n=1}^N x_n=\Sigma_{m=1}^M y_m=1$---that
restrict the agent states to lie in simplices become
$\Sigma_{n=1}^N e^{-\xi_n}=\Sigma_{m=1}^M e^{-\eta_m}=1$ in ${\bf \Xi}$. 

In this space the equations of motion become:
\begin{eqnarray} 
\dot{\xi}_i & = &
	-\beta_X [(Ae^{-\mbox{\boldmath $\eta$}})_i
	-e^{-\mbox{\boldmath $\xi$}}\cdot Ae^{-\mbox{\boldmath $\eta$}}]  
	- \alpha_X [\xi_i-e^{-\mbox{\boldmath $\xi$}}\cdot \mbox{\boldmath $\xi$}] ~,
\nonumber\\
\dot{\eta}_j & = &
	-\beta_Y [ (Be^{-\mbox{\boldmath $\xi$}})_j
	-e^{-\mbox{\boldmath $\eta$}}\cdot Be^{-\mbox{\boldmath $\xi$}}]  
	- \alpha_Y [\eta_j-e^{-\mbox{\boldmath $\eta$}}\cdot \mbox{{\boldmath $\eta$}}] ~,
	\nonumber\\
\label{InformationDynamics-Constant}
\end{eqnarray}
for $i = 1, \ldots, N$ and $j = 1, \ldots, M$
and where 
$e^{-\mbox{\boldmath $\xi$}}=(e^{-\xi_1},\ldots,e^{-\xi_N})$ and
$e^{-\mbox{\boldmath $\eta$}}=(e^{-\eta_1},\ldots,e^{-\eta_N})$. 

Recall that both the $\Delta R$ interaction term and the $\Delta H$
memory loss term are differences from means. This suggests yet another
transformation to remove these comparisons to the mean:
\begin{eqnarray} 
u_i & = & \xi_i - N^{-1} \sum_{n=1}^N \xi_n ~, \nonumber\\
v_j & = & \eta_j - M^{-1} \sum_{m=1}^M \eta_m ~, 
\label{CLR2}
\end{eqnarray}
with $i = 1, \ldots, N$ and $j = 1, \ldots, M$.
This leads to the normalized space in
${\bf R}^{N} \times {\bf R}^{M}$:
\begin{equation}
{\bf U}=({\bf u}, {\bf v})
  = (u_1, \ldots, u_{N}) \times (v_1, \ldots, v_{M}) ~, 
\end{equation}
with the constraints $\sum_{n=1}^N u_n = \sum_{m=1}^M v_m = 0$. ${\bf u}$
and ${\bf v}$ are the normalized self-informations relative to their means.
We refer to this space as \emph{information space}.

The combined coordinate transformation, Eq. (\ref{CLR2}) composed with Eq.
(\ref{CLR1}), gives the well known \emph{centered log-ratio}
coordinates \cite{Aitc86a}. The inverse transformation is:
\begin{eqnarray}
x_i & = & \frac{e^{-u_i}} {\sum_{n=1}^Ne^{-u_n}} ~, \nonumber\\
y_i & = & \frac{e^{-v_i}} {\sum_{m=1}^Me^{-v_m}} ~.
\end{eqnarray}

The resulting transformed adaptation equations directly model the
dynamics of uncertainties of agents' behavior:
\begin{eqnarray}
\dot{\bf u} & = &  -\beta_X
  \left[ A {\bf y} - \sum_{n=1}^N (A \mathbf{y})_n \right] 
  - \alpha_X {\bf u} ~, \nonumber \\ 
\dot{\bf v} & = & -\beta_Y
  \left[ B {\bf x} - \sum_{n=1}^N (B \mathbf{x})_n \right] 
  - \alpha_Y {\bf v} ~.
\label{EntropyEquations-Constant}
\end{eqnarray}
When the interaction matrices are normalized to zero mean, 
$\sum_{m=1}^M a_{im}=\sum_{n=1}^{N} b_{jn}=0$, the equations
simplify even further to
\begin{eqnarray}
\dot{\bf u} & = &  -\beta_X A {\bf y} - \alpha_X {\bf u} ~, \nonumber\\ 
\dot{\bf v} & = & -\beta_Y  B {\bf x} - \alpha_Y {\bf v} ~.
\label{EntropyEquations-Constant-Normalized}
\end{eqnarray}

The origin ${\bf O}=(0, 0, \ldots, 0)$ of the normalized information space
${\bf U}$ corresponds to random behavior:
$({\bf x}, {\bf y})=(1/N, \ldots, 1/N, 1/M, \ldots, 1/M)$. The Shannon entropy
of the choice distribution is maximized at
this point. In contrast, when agents choose an action with
probability $1$ the entropy vanishes and the agent state is located in
$\Delta$ at the simplex vertices and in ${\bf U}$ at infinity.

In Eqs. (\ref{EntropyEquations-Constant-Normalized}) the first term is related to 
information influx to an agent from outside; i.e., from other agents and
the environment. The second term is related to the information dissipation
due to internal memory loss. Eqs. (\ref{EntropyEquations-Constant-Normalized})
are useful for theory, for analysis in certain limits, as we will
shortly demonstrate, and for numerical stability during simulation,
which we will illustrate when considering example collectives below.
Note that Eqs. (\ref{LearningEquations-Constant}),
Eqs. (\ref{InformationDynamics-Constant}), and Eqs. 
(\ref{EntropyEquations-Constant}) are topologically orbit equivalent. 

\subsection{Self-organization Induced by Dynamics of Uncertainty} 

Equations (\ref{EntropyEquations-Constant}) describe a dynamics of uncertainty 
between deterministic and random behavior. Information influx occurs when the
agents adapt to environmental constraints and accordingly change their choice
distribution. Information dissipation occurs when memory loss dominates and
the agents increase their uncertainty to behave more randomly with less regard
to the environmental constraints. The dissipation rate $\gamma$ of the dynamics
in ${\bf U}$ is controlled entirely by the memory loss rate $\alpha$:
\begin{equation}
\gamma  =  \sum_{n=1}^N  \frac{\partial \dot{u}_n}{\partial u_n}
  + \sum_{m=1}^M \frac{\partial \dot{v}_m}{\partial v_m} 
  =  -N \alpha_X - M \alpha_Y ~.
\label{eq:diss}
\end{equation}
Therefore, Eqs. (\ref{EntropyEquations-Constant-Normalized}) are 
volume preserving in ${\bf U}$ when $\alpha_X = \alpha_Y = 0$.

In the case that agents behave without memory loss 
($\alpha_X=\alpha_Y=0$), if the interaction 
specified by $(A, B)$ is zero-sum, $B=-A^T$, 
and if, in addition, it determines an interior Nash equilibrium
$({\bf x}^*, {\bf y}^*)$ (see App. \ref{NashEquilibria}), then
the collective has a constant of motion:
\begin{equation}
E  =  \beta_X^{-1} D({\bf x}^{\ast}\parallel{\bf x}) + 
  \beta_Y^{-1} D({\bf y}^{\ast}\parallel{\bf y}) ~,
\label{ConstantofMotion}
\end{equation}
where $D({\bf p} \parallel {\bf q})=\Sigma_k p_k\log (p_k/q_k)$ is the
\emph{relative entropy} or the \emph{information gain} which measures
the similarity between probability distributions ${\bf p}$ and ${\bf q}$
\cite{Cove91}. (App. \ref{HamiltonianFormInformationSpace} 
gives the derivation of Eq. (\ref{ConstantofMotion}).) 
\begin{figure}[htbp]
 \begin{center}
  \leavevmode
  \includegraphics[scale=0.6]{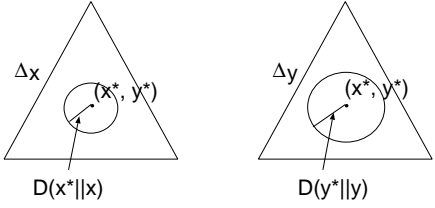}
 \end{center}
\caption{Dynamics of zero-sum interaction without memory loss: 
~Constant of motion 
$E  =  \beta_X^{-1} D({\bf x}^{\ast}\parallel{\bf x}) + 
  \beta_Y^{-1} D({\bf y}^{\ast}\parallel{\bf y})$ keeps the linear 
sum of distance between the interior Nash equilibrium 
and each agent's state. 
  }
\label{fig:RelativeEntropies}
\end{figure}
Since the constant of motion $E$ is a linear
sum of relative entropies, the collective maintains the information-theoretic 
distance between the interior Nash equilibrium 
and each agent's state. Thus, in the perfect memory case ($\alpha = 0$), by
the inequality $D({\bf p}\parallel{\bf q})\ge 0$, 
the interior Nash equilibrium 
cannot be reached unless the initial condition itself starts on it 
(Fig. \ref{fig:RelativeEntropies}). This is
an information-theoretic interpretation of the constant of motion noted in
Ref. \cite{Hof96}. 
Moreover, when $N=M$ the dynamics has a symplectic structure in ${\bf U}$
with the Hamiltonian $E$ given in Eq. (\ref{ConstantofMotion}) \cite{Hof96}.
In this case, Eqs. (\ref{EntropyEquations-Constant}) are described
quite simply,
\begin{equation}
\dot{\bf U}=J \nabla_{\bf U} E ~,
\end{equation}
with a Poisson structure $J$ ~
\begin{equation}
J = \left(\begin{array}{cc}
	O&P\\
	-P^T&O\\
\end{array}
\right) ~~\mbox{with}~~ P = -\beta_X \beta_Y A ~. 
\label{HamiltonianDynamics}
\end{equation}
Again, see App. \ref{HamiltonianFormInformationSpace}. 

When the bi-matrix interaction $(A, B)$ satisfies $B=A^T$, 
$E$ is a Lyapunov function of dynamics and decreases to $0$ 
over time \cite{Hof88}. In this case, each agents can 
adapt to environment independently and collective adaptation 
dynamics reach one of stable states. The Nash equilibria 
$({\bf x}^*, {\bf y}^*)$ may not be in the interior of 
the collective simplices $\Delta$. 
Note that symmetric neural networks have similar properties 
\cite{Hop82}. 

In some cases when neither $B=-A^T$ nor $B=A^T$, $E$ increases
non-monotonically, the dynamics in ${\bf U}$ diverges, and the Shannon
entropies of agents' choice distribution asymptotically decreases. (See
Figs. \ref{fig:hEntropy} and \ref{fig:dEntropy} below.) Note that in
single-agent adaptation with state $\bf x$ and normalizing the environment's
reinforcements to a probability distribution ${\bf p}_e$,
$D({\bf p}_e\parallel {\bf x})$ is always a Lyapunov function of the dynamics
and decreases monotonically. 
In mutual adaptation, however, agents adapt to a dynamic environment that
includes the other agents. As a result, in some cases, 
$E$, a linear sum of agent relative entropies, 
will itself exhibit nontrivial dynamics and, in addition, 
the uncertainties of agents' choices will asymptotically decrease. 

\begin{figure}[htbp]
 \begin{center}
  \leavevmode
  \includegraphics[scale=0.6]{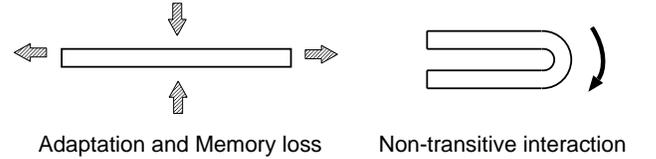}
 \end{center}
\caption{Schematic view of mutual adaptation: Effect of mutual 
  adaptation and memory loss produce unstable and stable directions. 
  The nontransitive structure of interactions leads to state-space folding.
  }
\label{fig:horseshoe}
\end{figure}

When agents adapt with memory loss ($\alpha$ $>0$), the dynamics is
dissipative. Since the memory loss terms induce information
dissipation, the dynamics varies between random and  
deterministic behavior in the information space. 
Notably, when the agents attempt to achieve this balance together by
interacting and, in particular, when the interaction has \emph{nontransitive}
structure, the dynamics can persistently wander in a bounded area in
information space. Since, in some cases, mutual adaptation and memory loss
produce successive stretching and folding, deterministic chaos can occur
with a significant range of $\alpha$, even with only two agents.
A schematic view of the flow in mutual adaptation is given in Fig. 
\ref{fig:horseshoe}. 

In the case that the agents are completely decoupled (or,  
in the case that $B=A^T$ and $\alpha_X=\alpha_Y=0$ for two agents),  
information space locally splits into subspaces governed by 
effects of mutual adaptation (information influx) and memory 
loss (information dissipation). They correspond to unstable 
and stable flow directions as in single agent adaptation. 
However, in the case that agents are coupled via nontransitive
interaction, mutual adaptation and memory loss affects with each other 
and horseshoe can be produced. 
Flow of information is multidimensional since each agent obtains information 
from its environment, organizes its behavior based on that
information, and that local adaptation is then fed back into the 
environment affecting other agents. 

In this case, ``weak'' uncertainty of behavior plays an important 
role in organizing the collective's behavior. Small fluctuations in
decision making can be amplified through repeated mutual adaptation with 
competitive interactions and dynamic memory stored in collectives
could exist shown by a positive metric entropy. 

Now consider many agents interacting. In the perfect memory case,
when the game is zero-sum and has an interior Nash equilibrium
$({\bf x}^{1*},{\bf x}^{2*}, \ldots, {\bf x}^{S*})$, following Eq.
(\ref{ConstantofMotion}), the following constant of motion exists: 
\begin{equation}
E   =  \sum_{s=1}^S \frac1\beta_s D({\bf x}^{s*}\parallel{\bf x}^s)
    =  \sum_{s=1}^S 
   \frac1\beta_s \left(\sum_{n^s=1}^{N^s} x_{n^s}^{s*} \log \frac{x_{n^s}^{s*}}{x_{n^s}}
    \right) ~. 
\label{MultiConstantOfMotion}
\end{equation}
Although, strictly speaking, Hamiltonian dynamics and the associated
symplectic structure of information space occurs only for two agents, one
can describe multiple agent dynamics as a generalized Hamiltonian system
\cite{Per90}. In the general case with $\alpha >0$, dissipative dynamics and
high-dimensional chaotic flows can give rise to several unstable directions,
since information influx has a network structure relative to the other agents.
At least $S$ stable directions are expected since memory loss comes from each
individual's internal dynamics. 

Summarizing, in single-agent adaptation, information flows unidirectionally
from the environment to the agent and the agent adapts its behavior to the
environmental constraints. Adaptation leads to 
$D({\bf p}_e \parallel {\bf x})\rightarrow 0$. For mutual adaptation in an
agent collective, however, information flow is multidimensional since 
each agent obtains information from its environment that includes the
other agents. In this situation, $E$ need not be a Lyapunov 
function for the dynamics. As we will see, when the dynamics is chaotic,
global information maximization is of doubtful utility and 
a dynamic view of adaptation shown in Fig. \ref{fig:horseshoe} 
is more appropriate. When dynamic memory in collectives emerges, 
collective adaptation becomes a non-trivial problem. 
A detailed dynamical and information theoretic analysis 
along these lines will be reported elsewhere. 

In the next section, we will give several  phenomenological examples
that captures collective adaptation. 

\section{Examples}
\label{Sec:Examples}

To illustrate collective adaptation, we now give several examples of the
dynamics in a static environment with two and three agents interacting via
versions of Matching Pennies and Rock-Scissors-Paper, games with 
non-transitive structures. 
App. \ref{ReinforcementSchemesInteractionMatrices} gives the 
details of the reinforcement schemes for these cases. The agents will
have equal adaptation rates ($\beta_X=\beta_Y=\cdots$) and the same number 
of actions ($N = M = L =\cdots$). In these simplified cases, the
equations of motion for two agents are given by 
\begin{eqnarray} 
\frac{\dot{x}_i}{x_i} & = & [ (A{\bf y})_i-{\bf x}\cdot A{\bf y}]
	+ \alpha_X [-\log x_i + \sum_{n=1}^N x_n\log x_n ] ~, \nonumber\\
\frac{\dot{y}_j}{y_j} & = & [ (B{\bf x})_j-{\bf y}\cdot B{\bf x}]
	+ \alpha_Y [-\log y_j + \sum_{m=1}^M y_m\log y_m] ~, \nonumber\\
\label{LearningEquations-Constant-Example}
\end{eqnarray} 
for $i, j = 1, \ldots, N$. 
A detailed analysis of this case with zero memory loss ($\alpha = 0$) 
is given in Ref. \cite{Hof88} in terms of asymmetric game dynamics. 
We will present results for zero and positive memory loss rates. 

We then consider three agents, for which the adaptation equations are 
\begin{eqnarray}
\frac{\dot{x}_i}{x_i} &=& [(A{\bf y}{\bf z})_i
	- {\bf x}\cdot A{\bf y}{\bf z}]
	+ \alpha_X [-\log x_i + \sum_{n=1}^N x_n\log x_n ] ~, \nonumber\\
\frac{\dot{y}_j}{y_j} &=& [(B{\bf z}{\bf x})_j
	- {\bf y}\cdot B{\bf z}{\bf x}]
	+ \alpha_Y [-\log y_j + \sum_{m=1}^M y_m\log y_m] ~, \nonumber\\
\frac{\dot{z}_k}{z_k} &=& [(C{\bf x}{\bf y})_k 
	- {\bf z}\cdot C{\bf x}{\bf y}]
	+ \alpha_Z [-\log z_k + \sum_{l=1}^L ~z_l \log z_l] ~, \nonumber\\
\label{LearningEquations-Constant-3Agents}
\end{eqnarray}
for $i, j, k = 1, \ldots, N$. 
We again will describe cases with and without memory loss.

Computer simulations are executed in the information space $\bf U$ 
and the results are shown in the state space $X$. We ignore the
dynamics on the boundary of the simplex and concentrate the case that 
all variables are greater than $0$ and less than $1$. 

\subsection{Two Agents Adapting under Matching Pennies Interaction}

In the matching pennies game, agents play one of two actions: heads ($H$)
or tail ($T$). Agent $X$ wins when the plays do not agree; agent $Y$ wins
when they do. Agent $X$'s state space is $\Delta_X = (x_1,x_2)$ 
with $x_i \in (0,1)$ and $x_1 + x_2 = 1$. That is, $x_1$ is the
probability that agent $X$ plays heads; $x_2$, tails. Agent $Y$
is described similarly. Thus, each agent's state space is effectively
one dimensional and the collective state space 
$\Delta  = \Delta_X \times \Delta_Y$, two dimensional. 

The environment for two agents interacting 
via the matching pennies game leads to the following matrices for 
Eqs. (\ref{LearningEquations-Constant-Example}): 
\begin{equation} 
A=\left[
  \begin{array}{cc}
    -\epsilon_X&\epsilon_X\\
    \epsilon_X &-\epsilon_X\\
  \end{array}
  \right] ~{\rm and}~
B=\left[
  \begin{array}{ccc}
    -\epsilon_Y&\epsilon_Y \\
    \epsilon_Y &-\epsilon_Y\\
  \end{array}
  \right] ~, 
\label{MPGame}
\end{equation}
where $\epsilon_X \in (0.0,1.0]$ and $-\epsilon_Y \in (0.0,1.0]$. 

\begin{figure}[htbp]
  \begin{center}
    \leavevmode
	\includegraphics[scale=0.53]{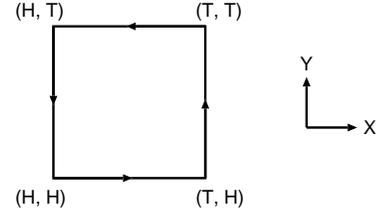}
  \end{center}
\caption{Flows on the boundary in Matching Pennies interaction: 
  Actions $H$ and $T$ correspond to ``heads'' and ``tails'', respectively.
  Arrows indicate the direction of adaptation dynamics on the boundary
  of the state space $\Delta$. 
  }
\label{fig:2MPInteraction}
\end{figure}

Figure \ref{fig:2MPInteraction} shows a heteroclinic cycle of adaptation
dynamics on the boundary of $\Delta$ when the $\alpha$s vanish. Flows on the
border occur only when agents completely ignore an action at the initial
state; that is, when $x_i(0)=0$ or $y_j(0)=0$ for at least one $i$ or $j$. 
Each vertex of the simplex is a saddle since the interaction 
is non-transitive. 

\begin{figure}[htbp]
  \begin{center}
    \leavevmode
	\includegraphics[scale=0.32]{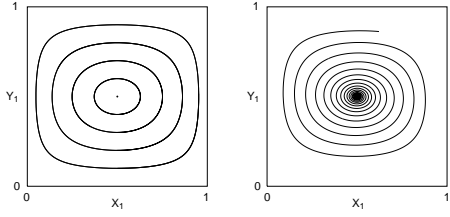}
  \end{center}
\caption{Adaptation dynamics in Matching Pennies interaction: Here $\epsilon_X = 0.5$ 
  and $\epsilon_Y = -0.3$ with (left) $\alpha_X = \alpha_Y = 0$ and (right)
  $\alpha_X = 0.02$ and $\alpha_Y = 0.01$. 
  }
\label{fig:2MPTrajectories}
\end{figure}

The Nash equilibrium $({\bf x}^*, {\bf y}^*)$ of the Matching Pennies 
game is in the center of $\Delta$: 
$({\bf x}^*, {\bf y}^*)=(\frac12, \frac12, \frac12, \frac12)$ 
and this is also a fixed point of the adaptation dynamics. 
The Jacobian at $({\bf x}^*, {\bf y}^*)$ is 
\begin{equation}
J=\left(
\begin{array}{cc}
-\frac{\alpha_X}{2}(1+\log 2) &-\frac{\epsilon_X}{2}\\
-\frac{\epsilon_Y}{2}&-\frac{\alpha_Y}{2}(1+\log 2)\\
\end{array}
\right)
\end{equation}
and its eigenvalues are
\begin{eqnarray}
\frac{4\lambda_i}{1+\log 2}  &=&  -(\alpha_X+\alpha_Y) \nonumber\\
   &\pm& \sqrt{(\alpha_X-\alpha_Y)^2+4\epsilon_X\epsilon_Y/(1+\log 2)^2} ~. 
\end{eqnarray}
In the perfect memory case ($\alpha_X=\alpha_Y=0$), trajectories 
near $({\bf x}^*, {\bf y}^*)$ are neutrally stable periodic orbits, since
$\lambda_i = \pm \frac12\sqrt{\epsilon_X\epsilon_Y}$ are pure imaginary. 
In the memory loss case ($\alpha_X > 0$ and $\alpha_Y>0$), 
$({\bf x}^*, {\bf y}^*)$ is globally asymptotically stable, since
Re($\lambda_1$) and Re($\lambda_2$) are 
strictly negative. Examples of the trajectories in 
these two cases are given in Figure \ref{fig:2MPTrajectories}. 

\subsection{Three Agents Adapting under Even-Odd Interaction}

Now consider extending Matching Pennies for two agents so that it determines
the interactions between three. Here we introduce the \emph{Even-Odd}
interaction in which 
there are again two actions, $H$ and $T$, but agents win according to whether
or not the number of heads in the group of three plays by the agents
is even or odd. The environment now is given by, for agent X,
\begin{equation}
a_{ijk}=\left\{
\begin{array}{ll}
\epsilon_X,  & \mbox{number of $H$s is even}\\
-\epsilon_X, & \mbox{otherwise}\\
\end{array}
\right. 
\end{equation}
with actions for agents $X$, $Y$, and $Z$ given by
$i, j, k = \{ H, T \}$ and $\epsilon_X \in (0.0, 1.0]$.
The interaction matrices $b_{jki}$ and $c_{kij}$ for agents $Y$
and $Z$, respectively, are given similarly, but with
$\epsilon_Y \in (0.0, 1.0]$ and $\epsilon_Z \in [-1.0,0.0)$. 
App. \ref{ReinforcementSchemesInteractionMatrices}
gives the details of the reinforcement scheme.

\begin{figure}[htbp]
  \begin{center}
    \leavevmode
	\includegraphics[scale=0.45]{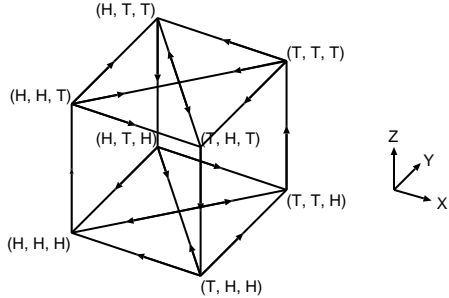}
  \end{center}
\caption{Flows on the state space boundary under the Even-Odd interactions: 
  $H$ and $T$ correspond to ``heads'' and ``tails'', respectively. Arrows
  indicate the direction of adaptation dynamics on $\Delta$'s boundary when
  the $\alpha$s vanish.
  }
\label{fig:3MPInteraction}
\end{figure}

Following the reasoning used in Matching Pennies, the collective state space 
$\Delta = \Delta_X \times \Delta_Y \times \Delta_Z$ is now a solid 
three-dimensional cube. Figure \ref{fig:3MPInteraction} shows a heteroclinic
network of adaptation dynamics on the boundary of $\Delta$ when $\alpha$s
vanish. Flows on $\Delta$'s boundary is shown in Fig. \ref{fig:3MPInteraction}. 

$\Delta$ is partitioned into four prism-shaped subspaces. Each
prism subspace has a heteroclinic cycle on the face that is also a
face of $\Delta$.

\begin{figure}[htbp]
  \begin{center}
    \leavevmode
	\includegraphics[scale=0.7]{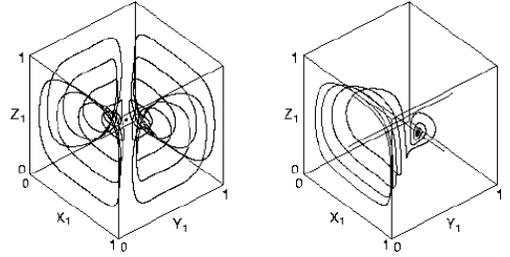}
  \end{center}
\caption{Dynamics of adaptation in the Even-odd interaction: $\epsilon_X = 0.5$,
  $\epsilon_Y = 0.2$, and $\epsilon_Z=-0.3$ with
  $\alpha_X = \alpha_Y = \alpha_Z=0$ in (left) and with
  $\alpha_X = \alpha_Y = 0$ and $\alpha_Z=0.01$ in (right). 
The trajectories with several initial conditions are shown in (left). 
  The neutral subspace is shown as the horizontal cross and 
  the trajectory chosen illustrates the attraction to a point in
  this subspace in (right).
  }
\label{fig:3MPTrajectories}
\end{figure}

The Nash equilibrium of the Even-Odd interaction is
$({\bf x}^*, {\bf y}^*, {\bf z}^*)=
(\frac12, \frac12, \frac12, \frac12, \frac12, \frac12)$ 
at the center of $\Delta$ and this is also 
a fixed point of the adaptation dynamics. 
The Jacobian there is 
\begin{equation}
J=\left(
\begin{array}{ccc}
-\alpha_X & 0 &0\\
0&-\alpha_Y&0\\
0&0&-\alpha_Z\\
\end{array}
\right) ~.
\end{equation}
Its eigenvalues are
$\lambda=-\alpha_X,-\alpha_Y,-\alpha_Z$. 
Thus, in complete memory case ($\alpha_X=\alpha_Y=\alpha_Z=0$), 
trajectories near $({\bf x}^*, {\bf y}^*, {\bf z}^*)$ 
are neutrally stable periodic orbits. 
With memory decay ($\alpha_X, \alpha_Y,\alpha_Z >0$), 
the $({\bf x}^*, {\bf y}^*, {\bf z}^*)$ is globally asymptotically stable.  
The hyperbolic fixed points in the top and bottom faces are unstable
in all cases. Examples of the trajectories are given 
in Figure \ref{fig:3MPTrajectories}. 

Notably, when a single agent (say, $Z$) has memory loss and others have
perfect memory, the crossed lines given by $\{z=x=0.5$, $z=y=0.5\}$
become an invariant subspace and trajectories are attracted to points
in this subspace. Thus, there are infinitely many neutrally stable points.

With $\alpha_X = \alpha_Y = 0$ and $\alpha_Z=0.01$, for example, the
adaptive dynamics alternates between a Matching Pennies interaction 
between agents $X$ and $Z$ by one between agents $Y$ and $Z$ during
the transient relaxation to a point on the invariant subspace. 

\subsection{Two Agents Adapting under Rock-Scissors-Paper Interaction}

In this subsection, we give an example of an environment in which agents 
have three actions. One of the most commonly studied games with three actions
is the Rock-Scissors-Paper (RSP) game, in which an agent playing Rock beats 
one playing Scissors, which in turn beats an agent playing Paper,
which finally beats Rock. 

First we examine two agents, which is a
straightforward implementation of the RSP game and then extend the
RSP interaction to three agents and analyze the higher-dimensional
behavior. The interaction matrices for these cases are given in
App. \ref{ReinforcementSchemesInteractionMatrices}. 

Under the RSP interaction each agent has the option of playing one
of three actions: ``rock'' (R), `scissors'' (S), and ``paper'' (P). 
Agent $X$'s probability of playing these are denoted $x_1$, $x_2$,
and $x_3$ and $x_1+x_2+x_3=1$.  
Agent $Y$ probabilities are given similarly. Thus, the
agent state spaces, $\Delta_X$ and $\Delta_Y$, are each two 
dimensional simplices, and the collective state space 
$\Delta = \Delta_X \times \Delta_Y$ is four dimensional. 

For two agents the environment is given by the interaction matrices
\begin{equation} 
A = \left[
  \begin{array}{ccc}
    \epsilon_X & 1 & -1\\
    -1 & \epsilon_X & 1\\
    1 & -1 & \epsilon_X\\
  \end{array}
  \right] ~{\rm and}~
B = \left[
  \begin{array}{ccc}
    \epsilon_Y & 1 & -1\\
    -1 & \epsilon_Y & 1\\
    1 & -1 & \epsilon_Y\\
  \end{array}
  \right] ~, 
\label{RSPGame}
\end{equation}
where $\epsilon_X, \epsilon_Y \in[-1.0,1.0]$ are the rewards for ties 
and normalized to 
\begin{equation}
A' = \left[
  \begin{array}{ccc}
    \frac23\epsilon_X & 1-\frac13\epsilon_X & -1-\frac13\epsilon_X\\
    -1-\frac13\epsilon_X & \frac23\epsilon_X & 1-\frac13\epsilon_X\\
    1-\frac13\epsilon_X & -1-\frac13\epsilon_X & \frac23\epsilon_X\\
  \end{array}
  \right]
\end{equation}
and
\begin{equation}
B' = \left[
  \begin{array}{ccc}
    \frac23\epsilon_Y & 1-\frac13\epsilon_Y & -1-\frac13\epsilon_Y\\
    -1-\frac13\epsilon_Y & \frac23\epsilon_Y & 1-\frac13\epsilon_Y\\
    1-\frac13\epsilon_Y & -1-\frac13\epsilon_Y & \frac23\epsilon_Y\\
  \end{array}
  \right]  ~.
\label{RSPGame_Normalized}
\end{equation}
Note that the reinforcements are normalized to zero mean and that this
does not affect the dynamics. 

\begin{figure}[htbp]
  \begin{center}
    \leavevmode
	\includegraphics[scale=0.5]{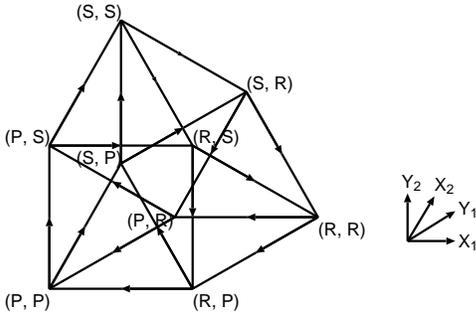}
  \end{center}
\caption{Flows on the boundary of the simplex in the Rock-Scissors-Paper
  interaction for two agents: $R$, $S$, and $P$ denote ``rock'', 
  ``scissors'', and ``paper'', respectively. The arrows indicate  the
  direction of the adaptation dynamics on the boundary of the collective
  state space $\Delta$ when the $\alpha$s vanish. 
  }
\label{fig:2RSPInteraction}
\end{figure}

The flow on $\Delta$'s boundary is shown in Fig. \ref{fig:2RSPInteraction}. 
This represents the heteroclinic network of adaptation dynamics on $\Delta$'s
edges when the $\alpha$s vanish. Each vertex is a saddle since the
interaction has non-transitive structure. 

The Nash equilibrium $({\bf x}^*,{\bf y}^*)$ is given by the
centers of the simplex: \begin{equation}
({\bf x}^*,{\bf y}^*)
 = (\frac{1}{3},\frac{1}{3},\frac{1}{3},
 \frac{1}{3},\frac{1}{3},\frac{1}{3}) ~. 
\end{equation}
This is also a fixed point of the adaptation dynamics. 
The Jacobian there is 
\begin{equation}
J=\left(
\begin{array}{cccc}
-\alpha_X & 0 &\frac{1+\epsilon_X}3&\frac23\\
0&-\alpha_X&-\frac23&\frac{-1+\epsilon_X}{3}\\
\frac{1+\epsilon_Y}{3}&\frac23&-\alpha_Y&0\\
-\frac23&\frac{-1+\epsilon_Y}{3}&0&-\alpha_Y\\
\end{array}
\right) ~.
\end{equation}
Its eigenvalues are
\begin{flushleft}
$2\lambda_i=-(\alpha_X+\alpha_Y)$
\end{flushleft}
\begin{equation}
\pm\sqrt{(\alpha_X-\alpha_Y)^2+\frac{4\left(\epsilon_X\epsilon_Y-3
\pm\sqrt{-3(\epsilon_X+\epsilon_Y)^2}\right)}{9}} ~.
\end{equation}
Thus, when $(A, B)$ is zero-sum ($\epsilon_X+\epsilon_Y=0$) and agents have
complete memory ($\alpha_X=\alpha_Y=0$), trajectories near
$({\bf x}^*, {\bf y}^*)$ are neutrally stable periodic orbits since all
$\lambda$'s are pure imaginary. The dynamics is Hamiltonian in this case. 
With memory decay ($\alpha_X, \alpha_Y >0$), 
and $|\alpha_X-\alpha_Y|<\frac23(\epsilon_X^2+3)$, 
$({\bf x}^*, {\bf y}^*)$ is globally asymptotically stable. 

For the nonzero-sum case, we will give examples of dynamics with
$\epsilon_X=0.5$, $\epsilon_Y=-0.3$, $\alpha_Y=0.01$. In this case, 
when $\alpha_X>\alpha_c$, $({\bf x}^*, {\bf y}^*)$ is globally
asymptotically stable. At the point $\alpha_c\sim0.055008938$, 
period-doubling bifurcation occurs. The example of two agents adapting
in the Rock-Scissors-Paper interaction adaptation dynamics illustrates
various types of low-dimensional chaos. We now explore several cases.

\subsubsection{Hamiltonian Limit}

\begin{figure}[htbp]
  \begin{center}
    \leavevmode
	\includegraphics[scale=0.75]{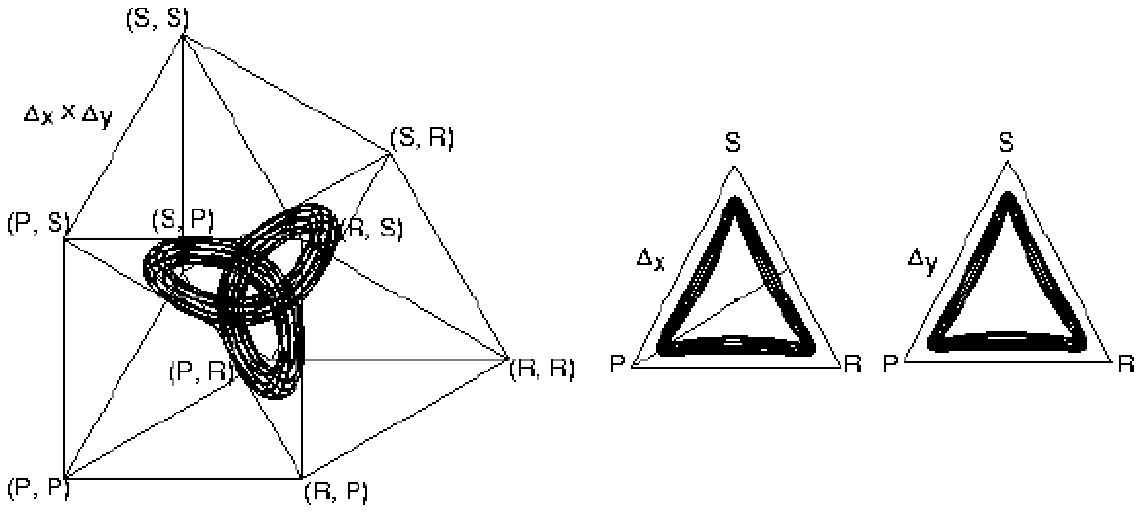}
	\includegraphics[scale=0.75]{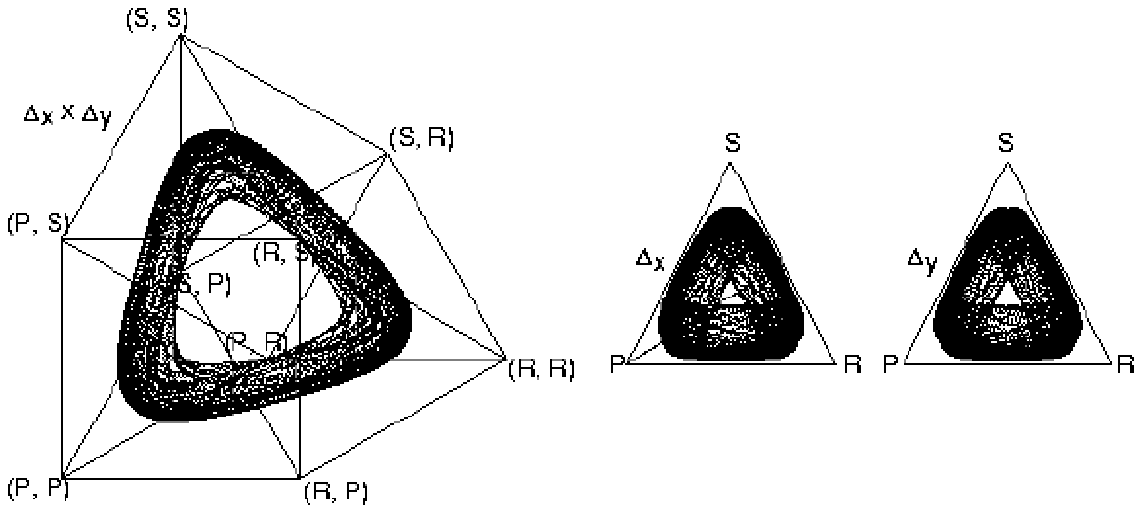}
  \end{center}
\caption{Quasiperiodic tori: Collective dynamics in $\Delta$ (left
  column) and individual dynamics projected onto $\Delta_X$ and
  $\Delta_Y$ respectively (right two columns). Here
  $\epsilon_X = - \epsilon_Y = 0.0$ and  $\alpha_X = \alpha_Y = 0$.
  The initial condition is (A): 
  $({\bf x},{\bf y}) = (0.26, 0.113333, 0.626667, 0.165, 0.772549,
  0.062451)$ for the top and (B): 
  $({\bf x},{\bf y}) = (0.05, 0.35, 0.6, 0.1, 0.2, 0.7)$ for the
  bottom. The constant of motion (Hamiltonian) is $E = 0.74446808 \equiv E_0 $. 
  The Poincar\'e section used for Fig. \ref{fig:HamilPSection} is given by
  $x_1=x_2$ and $y_1<y_2$ and is indicated here as the straight diagonal
  line in agent $X$'s simplex $\Delta_X$. 
  }
\label{fig:HamilIntegrable}
\end{figure}

\begin{figure}
  \begin{center}
    \leavevmode
	\includegraphics[scale=0.75]{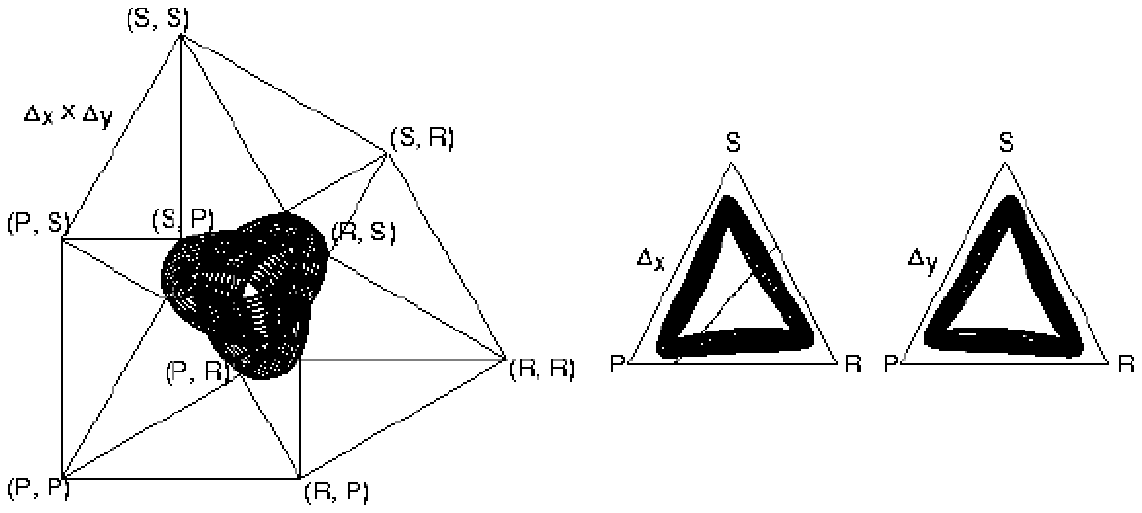}
	\includegraphics[scale=0.75]{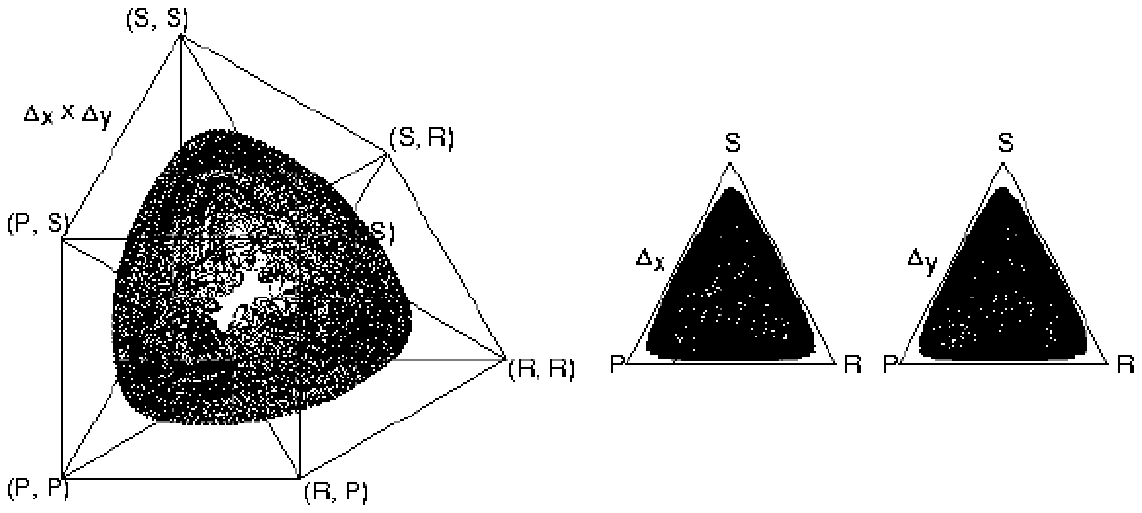}
  \end{center}
\caption{Quasiperiodic tori and chaos: Collective dynamics in $\Delta$
  (left column) and individual dynamics projected onto $\Delta_X$ and
  $\Delta_Y$, respectively (right two columns). Here
  $\epsilon_X = - \epsilon_Y = 0.5$ and $\alpha_X = \alpha_Y = 0$.  
  The initial conditions are the same as in Fig. \ref{fig:HamilIntegrable},
  (A) for top row and (B) for bottom rows, respectively. 
  Also, the constant of motion is the same: $E = E_0$. The Poincar\'e
  section is given by $3x_1-x_2-2/3=0$ and $y_1-3y_2+2/3<0$ and this
  is indicated as a straight line in $\Delta_X$.
  }
\label{fig:HamilChaos}
\end{figure}

When the agent memories are perfect ($\alpha_X=\alpha_Y=0$) and the
game is zero-sum ($\epsilon_X=-\epsilon_Y$), the dynamics in the 
information space $\bf U$ is Hamiltonian with a function consists 
of relative entropy 
$E=D({\bf x}^*\parallel {\bf x})+D({\bf y}^*\parallel{\bf y})$. 
The left columns of Figs. \ref{fig:HamilIntegrable} and 
\ref{fig:HamilChaos} give trajectories in the collective state
space $\Delta$, while the plots given in the middle and right columns
are these trajectories projected onto the individual agent
simplices, $\Delta_X$ and $\Delta_Y$. 
The trajectories were generated using a $4$th-order symplectic
integrator \cite{Yos90} in $\bf U$. 

When $\epsilon_X = -\epsilon_Y = 0.0$ it appears that 
the dynamics is integrable 
since only quasiperiodic tori exist for almost all initial conditions 
in our computer simulation. 
With some initial conditions, the tori is knotted to form trefoil. Otherwise, 
when $\epsilon_X = -\epsilon_Y > 0.0$, Hamiltonian chaos occurs with
positive-negative pairs of Lyapunov exponents. (See Table
\ref{Table:HamiLyap}.) The game-theoretic behavior of this example was
investigated briefly in Ref. \cite{Sat02}. The dynamics is very rich.
For example, there are infinitely many distinct behaviors 
near the fixed point at the center---the interior Nash equilibrium---and a
periodic orbit arbitrarily close to any chaotic one. 

\begin{figure}[htbp]
  \begin{center}
    \leavevmode
	\includegraphics[scale=0.75]{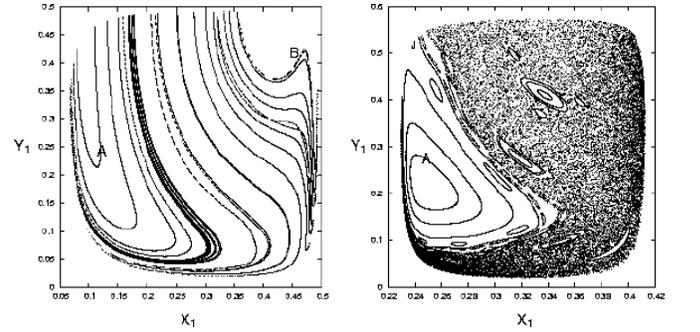}
  \end{center}
\caption{Poincar\'e sections of the behavior in the preceding two figures.
  That is, $\epsilon_X = - \epsilon_Y = 0.0$
  (left) and $\epsilon_X = - \epsilon_Y = 0.5$ (right). The Poincar\'e
  section is given by $x_1=x_2$ and $y_1<y_2$ (left) and
  $3x_1-x_2-2/3=0$ and $y_1-3y_2+2/3<0$ (right). There are 25 randomly
  selected initial conditions, including the two, (A) and (B), 
used in Figs.   \ref{fig:HamilIntegrable} and \ref{fig:HamilChaos}. The constant of
  motion ($E =E_0$) forms the outer border of the
  Poincar\'e sections. 
  }
\label{fig:HamilPSection}
\end{figure}

A more detailed view of the complex dynamics is given in Figure
\ref{fig:HamilPSection} which shows Poincar\'e sections of Eqs. 
(\ref{LearningEquations-Constant-Example})'s trajectories. 
The Poincar\'e section is
given by $\dot{u}_3 > 0$ and $\dot{v}_3 = 0$. In $({\bf x},{\bf y})$
space the section is determined by the constraints:
\begin{eqnarray}
(1 - \epsilon_X) y_1 & - & (1+\epsilon_X)y_2 +\frac23\epsilon_X< 0 ~,
	\nonumber\\
(1 - \epsilon_Y) x_1 & - & (1 + \epsilon_Y) x_2 + \frac23\epsilon_Y = 0
 ~.
\end{eqnarray}
These sections are indicated as the straight lines drawn in the $\Delta_X$
simplices of Figs. \ref{fig:HamilIntegrable} and \ref{fig:HamilChaos}. In
Figure \ref{fig:HamilPSection}, when $\epsilon_X=-\epsilon_Y=0.0$, closed
loops depending on the initial conditions exhibits tori in the Poincar\'e
section. When $\epsilon_X=-\epsilon_Y=0.5$, some tori collapse and become
chaotic. The scatter of dots among the remaining closed loop shows
characteristic Hamiltonian chaos. 

Table \ref{Table:HamiLyap} shows Lyapunov spectra in ${\bf U}$ for
dynamics with $\epsilon_X=-\epsilon_Y=0.0$ and $\epsilon_X=-\epsilon_Y=0.5$ 
with initial condition 
$({\bf x}(0), {\bf y}(0))=(x_1, 0.35, 0.65-x_1, 0.1, y_2, 0.9-y_2)$ 
with $E=E_0=0.74446808$ fixed. $(x_1, y_2)$ satisfies 
\begin{equation}
\frac{e^{-3(E_0+2\log3)}}{0.035}
= x_1(0.65-x_1)y_2(0.9-y_2). 
\end{equation}
When $x_1(0)=0.05$, the initial condition is (B):
$({\bf x}, {\bf y})=(0.05, 0.35, 0.6, 0.1, 0.2, 0.7)$, 
which we gave in the preceding examples. 
When $\epsilon_X=0.5$, the Lyapunov exponents indicate positive-negative
pairs for $x_1(0)=0.05, 0.06$ and $0.08$, which clearly show Hamiltonian
chaos. Note that $\lambda_2\simeq 0.0$, $\lambda_3\simeq 0.0$, and
$\lambda_4\simeq -\lambda_1$, as expected. 

\begin{table}[htbp]
\begin{center}
\begin{tabular}{@{}l@{}}
\noalign{\hrule height0.8pt}
\begin{tabular}{c|c|rrrrrr}
$\epsilon_X$&$\lambda$&$x_1(0)$=0.05&0.06&0.07&0.08&0.09&0.10 \\
\hline
      &$\lambda_1$& $+0.881$ & $+0.551$ & $+0.563$ & $+0.573$ & $+0.575$& $+0.589$\\
$0.0$ &$\lambda_2$& $+0.436$ & $+0.447$ & $+0.464$ & $+0.467$ & $+0.460$& $+0.461$ \\
      &$\lambda_3$& $-0.436$ & $-0.447$ & $-0.464$ & $-0.467$ & $-0.460$& $-0.461$\\
      &$\lambda_4$& $-0.881$ & $-0.551$ & $-0.563$ & $-0.573$ & $-0.575$& $-0.589$\\
\hline
    &$\lambda_1$& ${\bf +36.4}$  & ${\bf +41.5}$ & $+0.487$  &
${\bf +26.3}$&$+0.575$ & $+0.487$ \\
$0.5$ &$\lambda_2$& $+0.543$  & $+0.666$ & $+0.204$  & $+0.350$ & $+0.460$ & $+0.460$ \\
   &$\lambda_3$& $-0.637$  & $-0.666$ & $-0.197$  & $-0.338$ & $-0.460$ & $-0.467$ \\
   &$\lambda_4$& ${\bf -36.3}$  & ${\bf -41.5}$ & $-0.494$  &${\bf -26.3}$ & $-0.575$ & $-0.480$ \\
\hline
\end{tabular}\\
\noalign{\hrule height0.8pt}
\end{tabular}
\end{center}
\caption{Lyapunov spectra for different initial conditions (columns)
  and different values of the tie breaking parameter $\epsilon_X$. 
  The initial conditions are $(x_1, x_2, x_3, y_1, y_2, y_3) 
=(x_1, 0.35, 0.65-x_1, 0.1, y_2, 0.9-y_2)$
  with $E=E_0=0.74446808$ fixed. We choose the initial conditions 
$(x_1, y_2)$ = $(0.05, 0.2)$, 
$(0.06, 0.160421)$, $(0.07, 0.135275)$, 
  $(0.08, 0.117743)$, $(0.09, 0.104795)$, $(0.10, 0.0948432)$. 
The Lyapunov exponents are multiplied by
  $10^3$.  Note that $\lambda_2\simeq 0.0$, $\lambda_3\simeq 0.0$ and
  $\lambda_4\simeq -\lambda_1$ as expected. The Lyapunov exponents
  indicating chaos are shown in boldface. 
  }
\label{Table:HamiLyap}
\end{table}

\subsubsection{Conservative Dynamics}

\begin{figure}[htbp]
\begin{center}
	\leavevmode
	\includegraphics[scale=0.75]{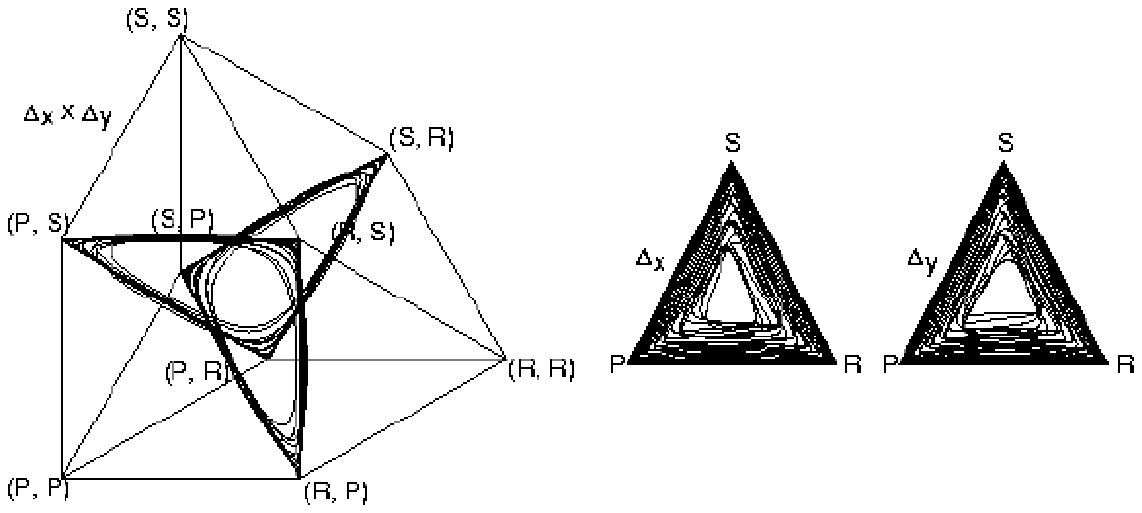}
	\includegraphics[scale=0.75]{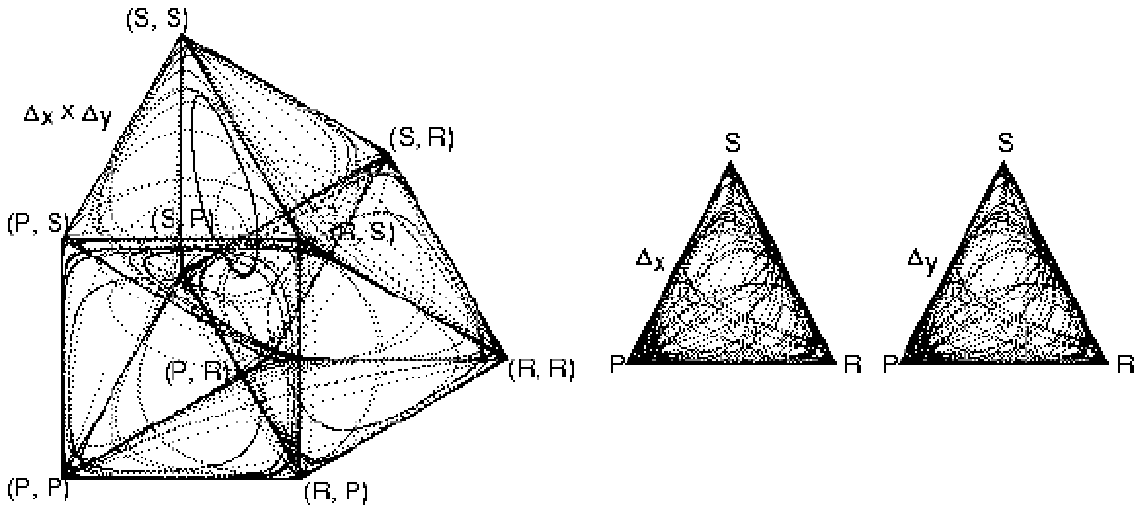}
\end{center}
\caption{Heteroclinic cycle with $\epsilon_X=-0.1$ and
  $\epsilon_Y = 0.05$ (top row). Chaotic transient to a heteroclinic
  network (bottom row) with $\epsilon_X=0.1$ and $\epsilon_Y = -0.05$).
  For both $\alpha_X = \alpha_Y = 0$. 
  }
\label{fig:HeteroClinic}
\end{figure}

With perfect memory ($\alpha_X=\alpha_Y=0$) and a game that is not zero-sum
($\epsilon_X \neq -\epsilon_Y$) the dynamics is conservative in 
$\bf U$ and one observes transients that are attracted
to heteroclinic networks in the state space $X$. (See Fig.
\ref{fig:HeteroClinic}.)

\begin{figure}[htbp]
\begin{center}
  \leavevmode
  \includegraphics[scale=0.9]{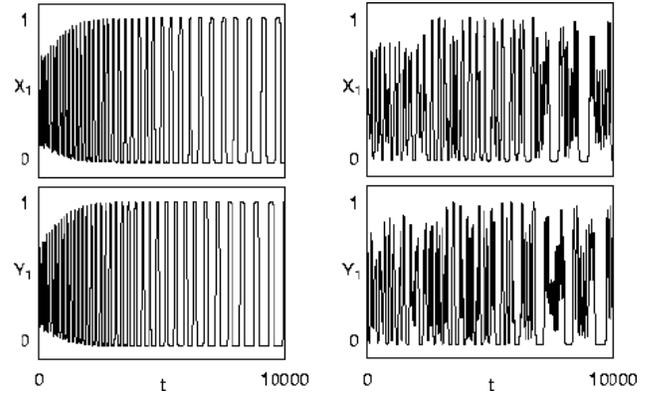}\hspace{5mm}
\end{center}
\caption{Time series of action probabilities during the heteroclinic
  cycles of Fig. \ref{fig:HeteroClinic}. $\epsilon_X=-0.1$ and 
  $\epsilon_Y = 0.05$ for the left column.
  The right column shows the chaotic transient to a possible
  heteroclinic cycles when $\epsilon_X=0.1$ and $\epsilon_Y = -0.05$.
  For both $\alpha_X = \alpha_Y = 0$.
  }
\label{fig:xHeteroClinic}
\end{figure}

\begin{figure}[htbp]
\begin{center}
  \leavevmode
  \includegraphics[scale=0.9]{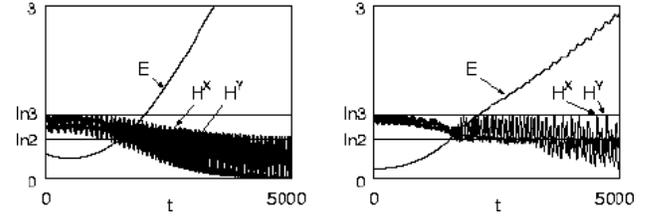}\hspace{5mm}
\end{center}
\caption{Dynamics of $H^X$, $H^Y$ and $E$ in conservative adaptive dynamics: 
  $\epsilon_X=-0.1$ and  $\epsilon_Y = 0.05$ for the left plot and
  $\epsilon_X=0.1$ and $\epsilon_Y = -0.05$ for the right. 
  For both $\alpha_X = \alpha_Y = 0$. Note that $E$ increases asymptotically 
  and $H^X$ and $H^Y$ tend to decrease.
  }
\label{fig:hEntropy}
\end{figure}

When $\epsilon_X+\epsilon_Y<0$, the behavior is intermittent and orbits
are guided by the flow on $\Delta$'s edges, which describes a network
of possible heteroclinic cycles. Since action ties are not rewarded there
is only one such cycle. It is shown in the top row of Fig. 
(\ref{fig:HeteroClinic}): 
$(R,P) \rightarrow (S,P) \rightarrow (S,R) \rightarrow (P,R)
\rightarrow (P,S) \rightarrow (R,S) \rightarrow (R,P)$. 
Note that during the cycle each agent switches between almost 
deterministic actions in the order 
$R \rightarrow S \rightarrow P$. The agents are out of 
phase with respect to each other and they alternate winning each
turn. 

With $\epsilon_X+\epsilon_Y>0$, however, the orbit is an infinitely
persistent chaotic transient \cite{Cha95}. Since, in this case, agent
$X$ can choose a tie, the cycles are not closed. For example, with 
$\epsilon_X > 0$, at $(R,P)$, $X$ has the option of moving to $(P,P)$ 
instead of $(S,P)$ with a positive probability. This embeds an 
instability along the heteroclinic cycle and so orbits are chaotic. 
(See Fig. \ref{fig:HeteroClinic}, bottom row.)

Figure \ref{fig:xHeteroClinic} shows the time series for these behaviors. 
Usually, in transient relaxation to heteroclinic cycle, the duration over
which orbits stay near saddle vertices increases exponentially. However,
for our case, it appears to increase subexponentially. This is because of
the very small exponent; $(1+\delta)^n\sim 1+n\delta+\ldots$ ~$(\delta<<1)$. 
In the second chaotic transient case, it still increases subexponentially, 
but the visited vertices change irregularly. 

Figure \ref{fig:hEntropy} shows the behavior of $H^X$, $H^Y$, and $E$. 
For both cases $E$ eventually increases monotonically and $H^X$ and
$H^Y$ asymptotically decrease. The agents show a tendency to decrease
choice uncertainty and to switch between almost deterministic actions. 
$H^X$ and $H^Y$ oscillate over the range $[0, \log 2]$ for $\epsilon_X=-0.1$
and  $\epsilon_Y = 0.05$ and over $[0, \log 3]$ for $\epsilon_X=0.1$
and $\epsilon_Y = -0.05$.

\subsubsection{Dissipative Dynamics}

\begin{figure}[htbp]
\begin{center}
	\leavevmode
	\includegraphics[scale=0.75]{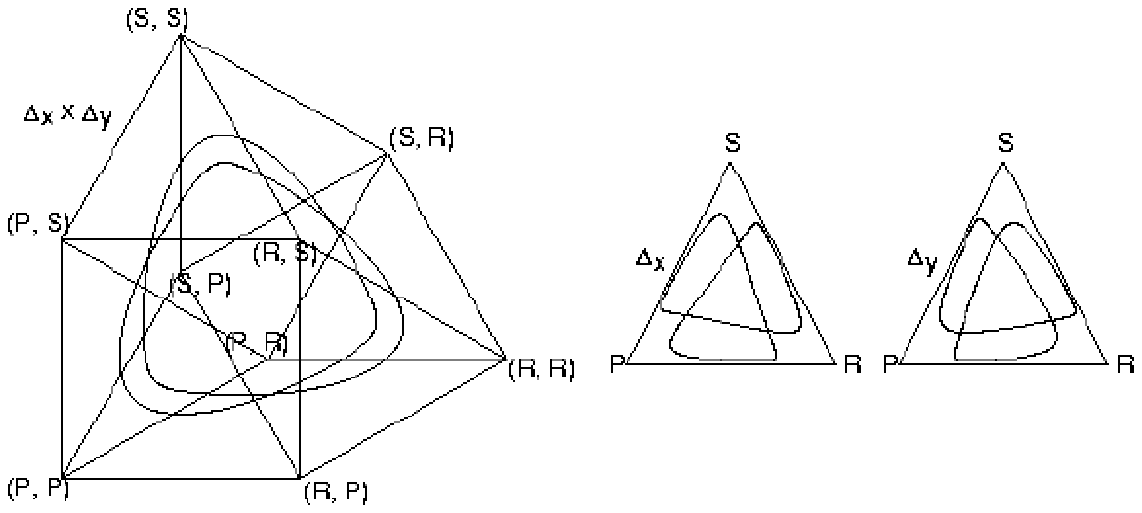}
	\includegraphics[scale=0.75]{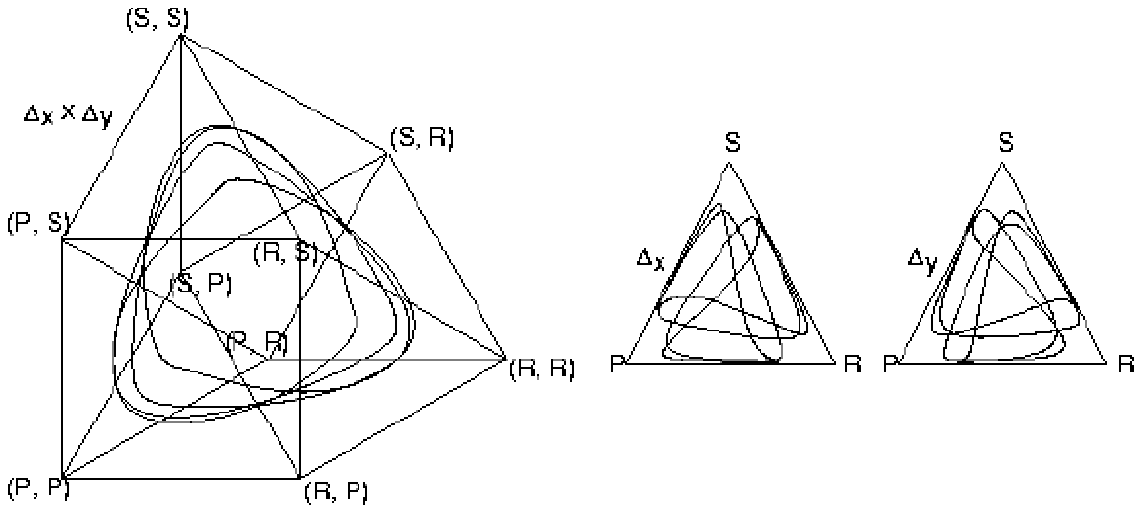}
	\includegraphics[scale=0.75]{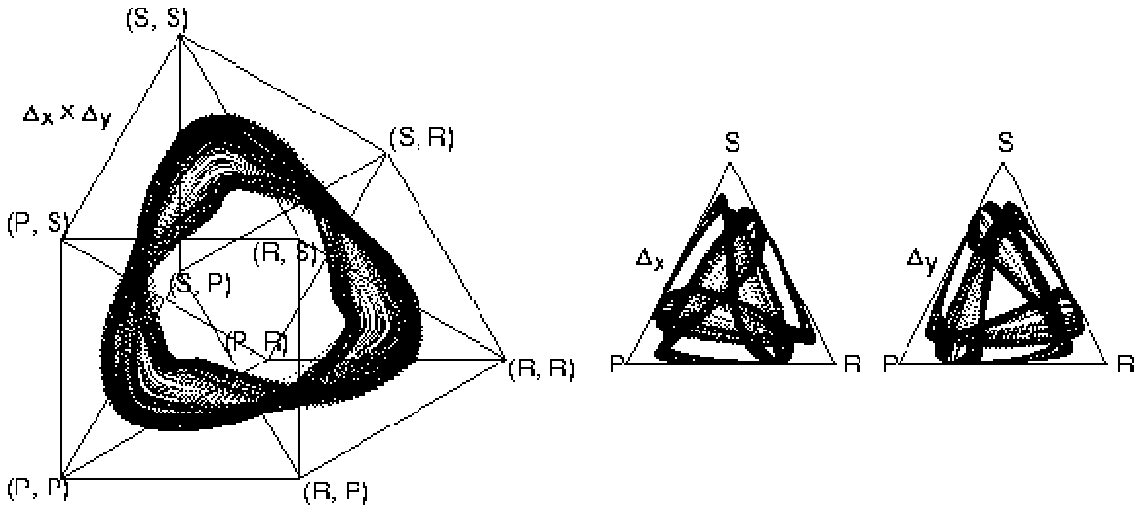}
\end{center}
\caption{Dissipative adaptive dynamics: Stable limit cycle for $\alpha_X = 0.025$
  (top),  $\alpha_X = 0.021$ (middle) and chaotic attractors with
  $\alpha_X = 0.0198$ (bottom). All cases have  $\epsilon_X = 0.5$,
  $\epsilon_Y=-0.3$ and $\alpha_Y = 0.01$. Period-doubling bifurcation
  to chaos occurs with decreasing $\alpha_X$.
  }
\label{fig:DissipativeChaos}
\end{figure}

If the memory loss rates ($\alpha_X$ and $\alpha_Y$) are positive, the
dynamics becomes dissipative in information space $\bf U$ and
exhibits limit cycles and chaotic attractors. (See Fig.
\ref{fig:DissipativeChaos}.) 

\begin{figure}[htbp]
\begin{center}
\leavevmode
	\includegraphics[scale=1.2]{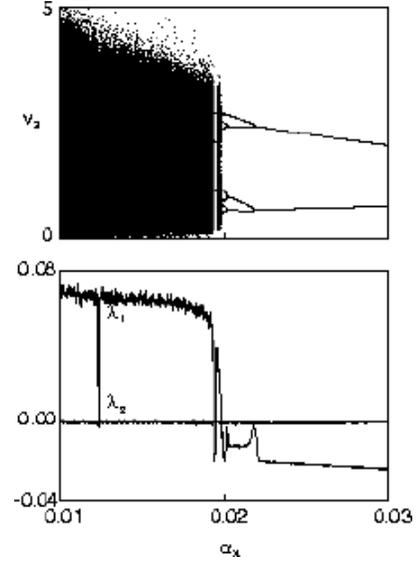}
\end{center}
\caption{Bifurcation diagram (top) of dissipative 
  dynamics (adapting with memory loss)
projected onto coordinate $v_3$ from the Poincar\'e section
  ($\dot{u}_3>0$, $\dot{v}_3=0$) and the largest two Lyapunov 
  exponents $\lambda_1$ and $\lambda_2$ (bottom) as a function of
  $\alpha_Y \in [0.01,0.03]$. Here with $\epsilon_X = 0.5$,
  $\epsilon_Y=-0.3$ and $\alpha_Y = 0.01$. 
  Simulations show that $\lambda_3$ and $\lambda_4$ are always
  negative. 
  }
\label{fig:BifnLCE}
\end{figure}

\begin{figure}[htbp]
\begin{center}
  \leavevmode
  \includegraphics[scale=0.9]{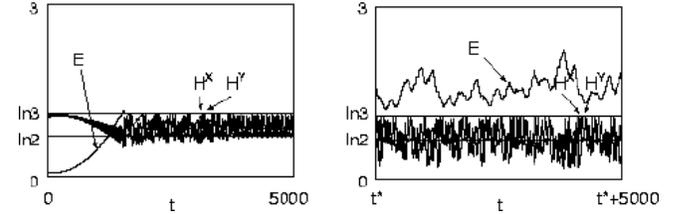}\hspace{5mm}
\end{center}
\caption{Dynamics of $H^X$, $H^Y$, and $E$ in dissipative adaptive dynamics: 
  $\epsilon_X=0.5$, $\epsilon_Y = -0.3$, and  $\alpha_Y = 0.01$ for both.
  $\alpha_X = 0.025$ for the left plot and $\alpha_X = 0.01$ for the right. 
  $t^* \approx 10^8$ in the right figure is the (rather long) transient time.  
  In both cases $E$ does not diverge due to memory loss.
  }
\label{fig:dEntropy}
\end{figure}

Figure \ref{fig:BifnLCE} (top) shows a diverse
range of bifurcations as a function of $\alpha_X$. It shows the dynamics
on the surface specified by $\dot{u}_3<0$ and $\dot{v}_3=0$ projected
onto $v_3$. The fixed point $({\bf x}^*, {\bf y}^*)$ becomes unstable 
when $\alpha_X$ is larger than $\alpha_c \approx 0.055008938$. Typically,
period-doubling bifurcation to chaos occurs with decreasing $\alpha_X$. 
Chaos can occur only when $\epsilon_X + \epsilon_Y > 0$ \cite{Sat03}. 

Figure \ref{fig:dEntropy} shows dynamics of $H^X$, $H^Y$, and $E$ in
dissipative adaptive dynamics. For both cases shown $E$ does not diverge 
due to memory loss. When $\alpha_X=0.025$, $H^X$ and $H^Y$ converge
to oscillations over the range $[\log 2, \log 3]$. When $\alpha_X=0.01$,
$H^X$ and $H^Y$ exhibit chaotic behavior over the range $[0, \log 3]$.

Figure \ref{fig:BifnLCE} (bottom) shows that the largest Lyapunov exponent
in ${\bf U}$ is positive across a significant fraction of the
parameter space; indicating that chaos is common. The dual aspects of
chaos, coherence and irregularity, imply that agents may behave
cooperatively or competitively (or switch between both). This ultimately
derives from agents' successive mutual adaptation and memory loss in
non-transitive interactions, such as in the RSP game; as was explained in
Sec. \ref{Sec:InfoSpace}. Note that such global behavior organization
is induced by each agents' self-interested and myopic adaptation and
``weak'' uncertainty of their environment.

\subsection{Three Agents Adapting under Rock-Scissors-Paper Interaction}

Consider three agents adapting via (an extension of) the RSP
interaction. Here the environment is given by the following interaction 

\begin{equation}
a_{ijk} = \left\{
  \begin{array}{ll}
	2 &~~~\mbox{Win over the others.}\\
	-2 &~~~\mbox{Lose to the other two.}\\
	1 &~~~\mbox{Win over one other.}\\
	-1 &~~~\mbox{Lose to one other.}\\
	\epsilon_X &~~~\mbox{Tie.}\\
  \end{array}
\right. 
\end{equation}
and similarly for $b_{jki}$ and $c_{kij}$, with $i, j, k = \{R, S, P\}$.
Here $\epsilon_X, \epsilon_Y, \epsilon_Z \in (-1.0, 1.0)$.
(See App. \ref{ReinforcementSchemesInteractionMatrices} for the
detailed listing of the reinforcement scheme.) 
As before we use normalized $a'_{ijk}$, $b'_{jki}$, and $c'_{kij}$:
\begin{equation}
a'_{ijk} = \left\{
  \begin{array}{ll}
	2-\frac{\epsilon_X}{5} &~~~\mbox{Win over the others.}\\
	-2-\frac{\epsilon_X}{5} &~~~\mbox{Lose to the other two.}\\
	1-\frac{\epsilon_X}{5}  &~~~\mbox{Win over one other.}\\
	-1-\frac{\epsilon_X}{5}  &~~~\mbox{Lose to one other.}\\
	\frac45{\epsilon_X} &~~~\mbox{Tie.}\\
  \end{array}
\right. 
\end{equation}
The normalization does not affect the dynamics. 

The Nash equilibrium $({\bf x}^*,{\bf y}^*, {\bf z}^*)$ is at the simplex
center:
\begin{equation}
({\bf x}^*,{\bf y}^*, {\bf z}^*)
 = (\frac{1}{3}, \frac{1}{3}, \frac{1}{3}, \frac{1}{3}, \frac{1}{3}, 
\frac{1}{3}, \frac{1}{3}, \frac{1}{3}, \frac{1}{3}) ~. 
\end{equation}
It is also a fixed point of the adaptation dynamics. 
The Jacobian there is 
\begin{equation}
J=\left(
\begin{array}{cccccc}
-\alpha_X & 0 &\frac13&\frac23&\frac13&\frac23\\
0&-\alpha_X&-\frac23&-\frac13&-\frac23&-\frac13\\
\frac13&\frac23&-\alpha_Y&0&\frac13&\frac23\\
-\frac23&-\frac13&0&-\alpha_Y&-\frac23&-\frac13\\
\frac13&\frac23&\frac13&\frac23&-\alpha_Z&0\\
-\frac23&-\frac13&-\frac23&-\frac13&0&-\alpha_Z\\
\end{array}
\right) ~.
\end{equation}
When $\alpha_X=\alpha_Y=\alpha_Z=\alpha$, its eigenvalues are
\begin{equation}
\lambda_i+\alpha =
 \frac{i}{\sqrt{3}} (-1,-1,-2,1,1,2) ~.
\end{equation}

\begin{figure}[htbp]
  \begin{center}
    \leavevmode
	\includegraphics[scale=0.4]{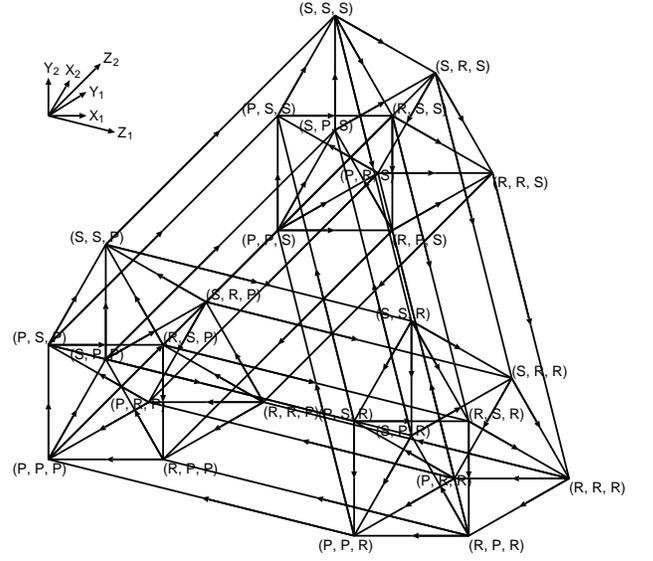}
  \end{center}
\caption{Flows on the simplex edges in three-agent RSP:
  Arrows indicate the direction of adaptation dynamics on
  $\Delta$'s boundary when the $\alpha$s vanish. 
  }
\label{fig:3RSPInteraction}
\end{figure}

\begin{figure}[htbp]
\begin{center}
	\leavevmode
	\includegraphics[scale=0.74]{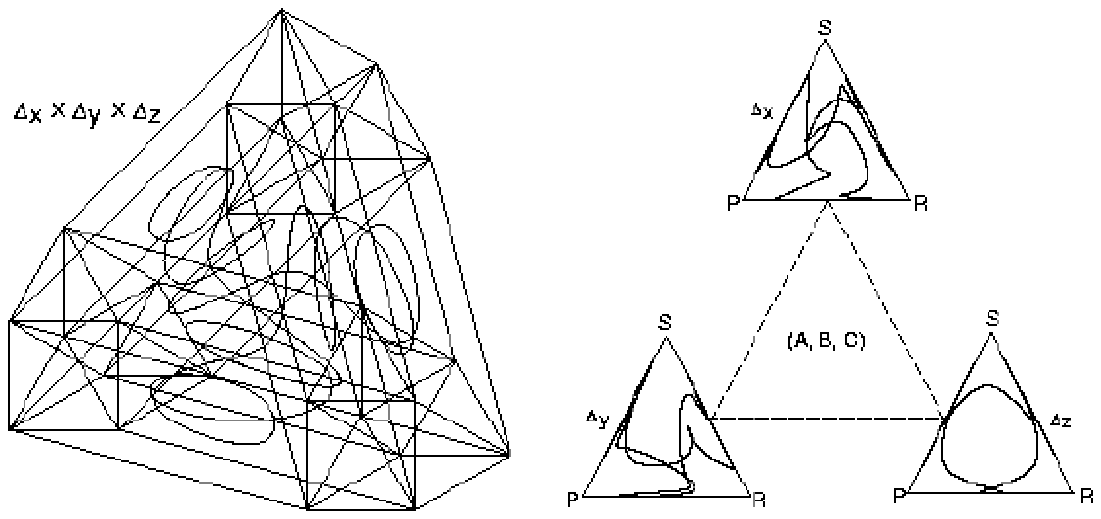}
	\includegraphics[scale=0.74]{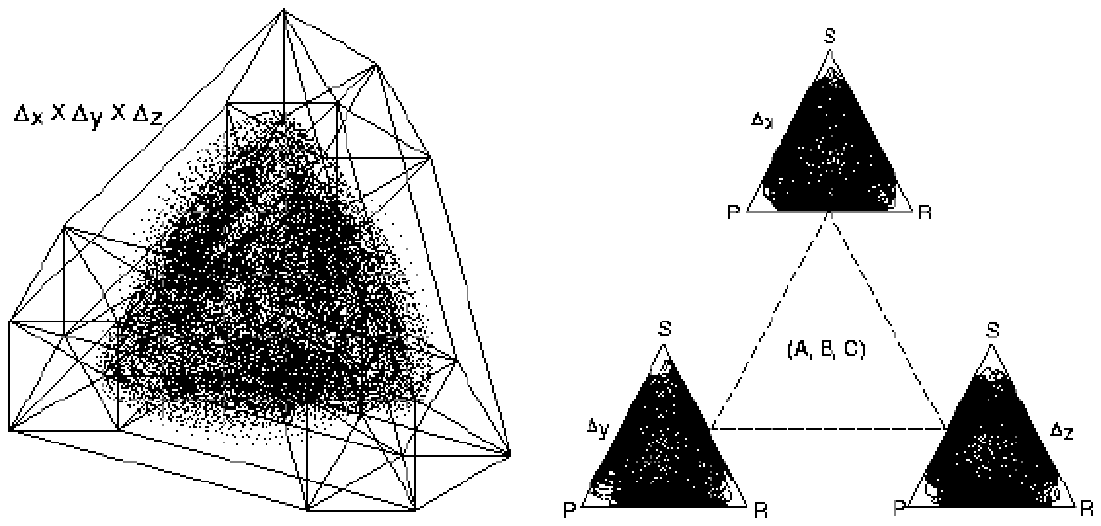}
\end{center}
\caption{Periodic orbit (top: 
$\epsilon_X = 0.5$, $\epsilon_Y=-0.365$,  $\epsilon_Z=0.8$) 
and chaotic orbit 
(bottom: $\epsilon_X = 0.5$, $\epsilon_Y=-0.3$,  $\epsilon_Z=0.6$) 
with the other parameters are $\alpha_X=\alpha_Y = \alpha_Z = 0.01$. 
The Lyapunov spectrum for chaotic dynamics is 
$(\lambda_1,\ldots,\lambda_6)=
(+45.2, +6.48, -0.336, -19.2, -38.5, -53.6)\times 10^{-3}$. 
  }
\label{fig:3PDissipativeChaos}
\end{figure}

In the perfect memory case ($\alpha_X=\alpha_Y=\alpha_Z=0$), trajectories 
near $({\bf x}^*, {\bf y}^*, {\bf z}^*)$ are neutrally stable 
periodic orbits, since the $\lambda$s are pure imaginary. 
In the memory loss case ($\alpha_X, \alpha_Y, \alpha_Z > 0$), 
$({\bf x}^*, {\bf y}^*, {\bf z}^*)$ is asymptotically stable, since
all Re($\lambda_i$) are strictly negative. 
One expects multiple attractors in this case. 

The collective state space $\Delta$ is now 6 dimensional, being the
product of three two-dimensional agent simplices 
$\Delta=\Delta_X\times\Delta_Y\times\Delta_Z$. The flow on
$\Delta$'s boundary is shown in Fig. \ref{fig:3RSPInteraction},
giving the adaptation dynamics on the edges of $\Delta$ 
when the $\alpha$s vanish. 

We give two examples with $\alpha_X=\alpha_Y=\alpha_Z=0.01$, 
$\epsilon_X=0.5$, $\epsilon_Y=-0.365$,  $\epsilon_Z = 0.8$ (top: limit cycle) 
and $\epsilon_X=0.5$, $\epsilon_Y=-0.3$,  $\epsilon_Z = 0.6$ (bottom: chaos)
in Fig. \ref{fig:3PDissipativeChaos}. 
Chaos is typically observed when $\epsilon_X+\epsilon_Y+\epsilon_Z >0$. 
Limit cycles are highly complex manifolds depending 
on the 6-dimensional heteroclinic network on the simplex boundary. 
The Lyapunov spectrum for the chaotic dynamics is 
$(\lambda_1, \ldots, \lambda_6)=( +45.2, +6.48,-0.336$, 
$-19.2, -38.5, -53.6)\times 10^{-3}$. The dynamics has two positive Lyapunov exponents.  Note that this dynamics could have many neutrally 
stable subspaces in three or more dimensions. These subspaces act as
quasistable attractors and may even have symplectic structure. These
properties of high-dimensional dynamics will be reported elsewhere.

\section{Concluding Remarks}
\label{Sec:TheEnd}

We developed a class of dynamical systems for collective adaptation.  
We started with very simple agents, 
whose adaptation was a dynamic balance between adaptation
to environmental constraints and memory loss. 
A macroscopic description of a network of adaptive agents was 
produced. In one special case we showed that the dynamical system reduces
to replicator equations, familiar in evolutionary game theory and 
population biology. In a more general setting, we investigated
several of the resulting periodic, intermittent, and chaotic behaviors
in which agent-agent interactions were explicitly given as game interactions. 

Self-organization induced by information flux was discussed using an
information-theoretic viewpoint. We pointed out that unlike single-agent
adaptation, information flow is multidimensional in collective
adaptation and that global information maximization is of doubtful 
utility and a dynamic view of adaptation is more appropriate. 
We also noted that only with two agents via nontransitive interactions, 
horseshoe in the information space can be produced due to 
the agents' local adaptation which amplifies fluctuations in behavior 
and to memory loss stabilizing behavior. 
Since deterministic chaos occurs even in this simple setting, one expects
that in higher-dimensional and heterogeneous adaptive systems intrinsic
unpredictability would become a dominant collective behavior. 
When dynamic memory stored in collectives emerges, 
collective adaptation becomes a non-trivial problem. 
A detailed information theoretic and dynamical systems 
theoretic analysis will be reported elsewhere. 

We close by indicating some future directions in which to extent the
model. 

First, as we alluded to during the development, there are
difficulties of scaling the model to large numbers of agents. We focused
on collectives with global coupling between all agents. However, in this
case, the complexity of interaction terms grows exponentially with number of
agents, which is both impractical from the viewpoints of analysis and
simulation, and unrealistic for natural systems that are large collectives.
The solution to this, given in App. \ref{NetworkInteractions}, 
is to develop either spatially 
distributed agents collectives or to extend the equations to include
explicit communication networks between agents. Both of these extensions
will be helpful in modeling the many adaptive collectives noted in the
introduction. 

Second, important for applications, is to develop the stochastic
generalization of the deterministic equations of motion which accounts for
the effects of finite and fluctuating numbers of agents and also
finite histories for adaptation. Each of these introduces its own kind 
of sampling stochasticity and will require a statistical dynamics 
analysis reminiscent of that found in population genetics \cite{Nimw97a}. 
It is also important to consider the effects of asynchrony of adaptive
behavior in this case. 

Third, one necessary and possibly difficult extension will be to agents that adapt
continuous-valued actions---say, learning the spatial location of
objects---to their environments. Mathematically, this requires a
continuous-space extension of the adaptation equations
(Eq. (\ref{LearningEquations})) and this results in models
that are described by PDEs \cite{Hofb97a}.

Finally, another direction, especially useful if one attempts to
quantify global function in large collectives, will be
structural and information-theoretic analyses 
of local and global adaptive behaviors \cite{Sha84,Cru89}. 
Analyzing the stored information and the causal architecture 
\cite{Crut98d,Crut01a} in each agent versus that in the collective, 
communication in networks, and emerging hierarchical 
structures in collective adaptation are projects 
now made possible using this framework. 


\begin{appendix}

\section{Continuous Time}
\label{ContinuousTimeLimits}

Here we give the derivation of the continuous-time limits that lead to
the differential equations from the original stochastic discrete-time
adaptation model.

Denote the agent-agent interaction time scale, number of interactions
per adaptation interval, and adaptation time scale as $d\tau$,
$T$, and $t$, respectively. We assume that adaptation is very slow
compared to agent-agent interactions and take the limits
$d\tau\rightarrow 0$ and $T\rightarrow\infty$, keeping $dt=T d\tau$
finite. Then we take the limit $dt\rightarrow 0$ to get the derivative
of the vector ${\bf Q}^X(t)$. 

With Eq. (\ref{MultiMemoryUpdate}) and $Q_i^X(0)=0$, we have 
\begin{equation}
Q_i^X(T)=\frac1T\sum_{k=1}^{T} \left[
\sum_{m=1}^{M}\delta_{im}(k) r_{im}^X(k) 
- \alpha_X Q_i^X (k)\right] ~.
\end{equation}
Thus, for continuous-time, when action $i$ is chosen by $X$ at step $t$,
\begin{eqnarray}
&&\frac{Q_i^X (t+dt) - Q_i^X (t)}{dt} \nonumber\\
	 &=&  \frac{1}{T dt}
	\sum_{k=Tt}^{T(t+dt)}\left[\sum_{m=1}^{M}
	\delta_{im}(\frac kT)r_{im}^X(\frac{k}{T}) 
	- \alpha_X Q_i^X (\frac{k}{T})\right] ~. \nonumber\\
\end{eqnarray}
Taking $T\rightarrow \infty$ and $d\tau\rightarrow 0$, we have
\begin{eqnarray}
&&\frac{Q_i^X(t+dt) - Q_i^X(t)}{dt} \nonumber\\
	  &=&  \frac{1}{dt} \int_{t}^{t+dt} \left[
\sum_{m=1}^{M}\delta_{im}(s)r_{im}^X(s)\right]ds \nonumber\\
    &-& \alpha_X\frac{1}{dt}\int_{t}^{t+dt}Q_i^X (s) ds ~. 
\label{LearningLimit}
\end{eqnarray}

Assuming $r_{ij}^X(t)$ changes as slowly as the adaptive dynamics, 
$r_{ij}^X(t)$ is constant during the adaptation interval $t\sim t+dt$. 
If we assume in addition that the behaviors of two agents
$X$ and $Y$ are statistically independent at time
$t$, then the law of the large numbers gives
\begin{eqnarray}
&&\frac{1}{dt}\int_{t}^{t+dt}
 \left[\sum_{m=1}^{M}\delta_{im}(s)r_{im}^X(s) 
\right]ds \nonumber\\
&\rightarrow& \sum_{m=1}^{M} r_{im}(t)y_m(t)\equiv R_i^X(t) ~.
\label{InteractionLimit}
\end{eqnarray}
Now take $dt\rightarrow 0$. Eqs. (\ref{LearningLimit}) and
(\ref{InteractionLimit}) together give
\begin{equation}
\dot{Q}_{i}^X(t)= R_i^X(t)-\alpha_X Q_{i}^X(t) ~,
\end{equation}
for the continuous-time updating of the reinforcement memory. When
environment is static given as $r_{ij}^X(t)=a_{ij}$, then
\begin{equation}
R_i^X(t) = \sum_{n=1}^N a_{in} y_i(t) ~. 
\end{equation}
The single-agent case is given by letting ${\bf y}=(1,0,0,\ldots,0)$
fixed and $a_{i1}=a_i$, $i = 1,\ldots,N$.

\section{Network interactions}
\label{NetworkInteractions}

\begin{figure}[htbp]
\begin{center}
\includegraphics[scale=0.4]{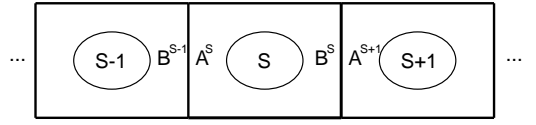}
\end{center}
\caption{Agent $s$ interacts with agent $s-1$ through bi-matrices
  $(A^s, B^{s-1})$ and agent $s+1$ through $(B^s, A^{s+1})$.
  }
\label{fig:lattice}
\end{figure}

We can describe heterogeneous network interactions within our model. 
We give an example of a model for lattice interactions here. Agents
$s=1, 2, \ldots, S$ are on a spatial lattice: agent $s$ interacts
with agent $s-1$ through bi-matrices $(A^s, B^{s-1})$ and agent
$s+1$ through $(B^s, A^{s+1})$. Each bi-matrix is $2\times 2$.
See Fig. \ref{fig:lattice}.

Agents choose actions among the $2 \times 2$ action pairs for both the
right and left neighboring agents. The action pairs are
$(1, 1), (1, 2), (2, 1), (2, 2)$ and are weighted with probabilities
$x_{1}, \ldots, x_{4}$. Inserting the interaction bi-matrices
into the S-agent adaptive dynamics of Eq. (\ref{MultiLearningEquations}) gives
\begin{eqnarray}
\frac{\dot{x^s_i}}{x^s_i}  
  &=&   \beta_s \left[(A^s {\bf x}^{s-1})_i - {\bf p}^s\cdot A^s{\bf x}^{s-1} \right.\nonumber\\
  &+& \left.(B^s {\bf x}^{s+1})_i - {\bf q}^s\cdot B^s{\bf x}^{s+1} \right] \nonumber\\
  &+&   \alpha_s (-\log x^s_i-\sum_{n=1}^{4} x^s_n \log x^s_n) ~, 
\label{LatticeLearningEquations}
\end{eqnarray}
where $\Sigma x^s_i=1$ and 
${\bf p}^s=(x^s_1+x^s_{2}, x^s_{3}+ x^s_{4})$,  
${\bf q}^s=(x^s_1+x^s_{3},  x^s_{2}+x^s_{4})$. 
In a similar way, arbitrary network interactions can be described by
our adaptive dynamics given in Eqs. (\ref{MultiLearningEquations}). 

\section{Nash Equilibria}
\label{NashEquilibria}

The \emph{Nash equilibria} $({\bf x}^*, {\bf y}^*)$ of the bi-matrix
game $(A, B)$ are those states in which all players can do no better by
changing state; that is,
\begin{equation}
{\bf x}^*A{\bf y}^*\ge{\bf x}A{\bf y}^* ~\mbox{and}~
	{\bf y}^*B{\bf x}^*\ge{\bf y}B{\bf x}^* ~,
\end{equation}
for all $({\bf x}, {\bf y}) \in \Delta_X \times \Delta_Y$.
If they exist in the interior, 
the solutions of the following simultaneous equations 
are Nash equilibria:
\begin{eqnarray}
&&(A{\bf y})_i  = 
	(A{\bf y})_1 ~\mbox{and}~ (B{\bf x})_j=(B{\bf x})_1 \nonumber\\
&&\Longleftrightarrow (A{\bf y})_i-{\bf x}A{\bf y}=(B{\bf x})_j-{\bf y}B{\bf x}=0 ~,
\label{NashEquations}
\end{eqnarray}
where
$\Sigma_{n=1}^N x_n = \Sigma_{m=1}^M y_m = 1$.

It is known that $N=M$ is a necessary condition for the existence of
a unique Nash equilibrium in the interior of $\Delta$. With $N=M$
in the perfect memory case ($\alpha_X = \alpha_Y = 0$), the unique
Nash equilibrium, if it exists, is the fixed point given by the
intersection of the $x$- and $y$-nullclines of Eqs.
(\ref{LearningEquations-Constant}).

This Nash equilibrium is not asymptotically stable, but the time
average of trajectories converges to it. To see this, suppose that
$x_i(t) > \delta$ for all $t$ sufficiently large, we have 
\begin{eqnarray}
\frac{d}{dt}(\log x_i) & = & \frac{\dot{x_i}}{x_i}
	= (A{\bf y})_i-{\bf x}A{\bf y} ~,\nonumber\\
\frac{d}{dt}(\log y_j) & = & \frac{\dot{y_j}}{y_j}
	= (B{\bf x})_j-{\bf y}B{\bf x} ~.
\end{eqnarray}
Integrating the both sides from $0$ to $T$ and dividing by $T$, we get
\begin{eqnarray}
\frac{\log x_i(T)-\log x_i(0)}{T} & = &
	\sum_{m=1}^M a_{im}\overline{y}_m -S_A ~, \nonumber\\
\frac{\log y_j(T)-\log y_j(0)}{T} & = &
	\sum_{n=1}^N b_{jn}\overline{x}_n -S_B ~,
\end{eqnarray}
where 
\begin{equation}
\overline{x}_i = T^{-1} \int_0^Tx_i dt ~\mbox{and}~ 
\overline{y}_j = T^{-1} \int_0^Ty_j dt ~,
\end{equation}
and
\begin{equation}
S_A = T^{-1} \int_0^T{\bf x}A{\bf y} dt ~\mbox{and}~ 
S_B = T^{-1} \int_0^T{\bf y}B{\bf x}dt ~.
\end{equation}
Letting $T\rightarrow\infty$, the left-hand sides converge to $0$.
Thus, $\overline{\bf x}$ and $\overline{\bf y}$ are a solution of Eqs. 
(\ref{NashEquations}). (This proof follows Ref. \cite{Sch81a}.) 

\section{Hamiltonian Dynamics}
\label{HamiltonianFormInformationSpace}

Consider a game $(A,B)$ that admits 
an interior Nash equilibrium 
$({\bf x}^*, {\bf y}^*) \in \Delta_X\times\Delta_Y$,  
and is zero-sum ($B=-A^T$), then
\begin{equation}
E = \beta_X^{-1} D({\bf x}^*\parallel {\bf x}) +
  \beta_Y^{-1} D({\bf y}^*\parallel{\bf y})
\end{equation}
is a constant of the motion. This follows by direct calculation:
\begin{eqnarray}
\frac{dE}{dt}
	& = & -\frac{1}{\beta_X}\sum_{n=1}^N x_n^*\frac{\dot{x}_n}{x_n}
		-\frac{1}{\beta_Y}\sum_{m=1}^M y_m^*\frac{\dot{y}_m}{y_m} \nonumber\\
	& = & -({\bf x}^*A{\bf y}-{\bf x}A{\bf y})
		-({\bf y}^*B{\bf x}-{\bf y}B{\bf x}) \nonumber\\
	& = & ({\bf x}^*-{\bf x})A({\bf y}^*-{\bf y})+({\bf y}^*-{\bf y})
		B({\bf x}^*-{\bf x}) \nonumber\\
    & = & 0 ~. 
\end{eqnarray}
This holds for any number of agents. Give the agents
equal numbers of actions ($N=M$) and set $\alpha$ to zero (perfect 
memory) and make all $\beta$s finite and positive. Then the adaptive
dynamics is Hamiltonian in the information space 
${\bf U}=({\bf u}, {\bf v})$  
with the above constant of motion $E$,  
\begin{equation}
\dot{\bf U}=J \nabla_{\bf U} E ~,
\end{equation}
with Poisson structure $J$, 
\begin{equation}
J=\left(\begin{array}{cc}
O&P\\
-P^T&O\\
\end{array}
\right) ~~\mbox{with}~~ P = -\beta_X \beta_Y A ~.
\label{PoissonStructure}
\end{equation}
\emph{Proof}:
\begin{eqnarray}
&&\frac{\partial E}{\partial u_i}
   =  \frac{\partial}{\partial u_i}
	\left[ \beta_X^{-1} \sum_{n=1}^N x_n^* \log x_n^* 
	+ \beta_Y^{-1} \sum_{n=1}^N y_n^* \log y_n^* \right. \nonumber\\
  &  & ~~~~~~\left. - \beta_X^{-1} \right( \sum_{n=1}^N x_n^* u_n
	- \log(\sum_{n=1}^N e^{-u_n}) \left) \right. \nonumber\\
  & &   ~~~~~~\left. - \beta_Y^{-1} \left( \sum_{n=1}^N y_n^* v_n
	- \log(\sum_{n=1}^N e^{-v_n}) \right) \right] \nonumber\\
  & & ~~~~~~=  \beta_X^{-1} (x_i^*-\frac{e^{-u_i}}{\sum_{n=1}^N e^{-u_n}})
   =  \beta_X^{-1} (x_i^*-x_i) ~, \\
&&\frac{\partial E}{\partial v_j} = \beta_Y^{-1} (y_j^*-y_j) ~. 
\end{eqnarray}
Since $({\bf x}^*, {\bf y}^*)$ is an interior Nash equilibrium,  
with Eq. (\ref{NormalRewards}), $(A{\bf y}^*)_i=(B{\bf x}^*)_j=0$.
Thus,
\begin{eqnarray}
A\frac{\partial E}{\partial {\bf v}} & = & -\frac1\beta_Y A{\bf y} ~,
	\nonumber\\
B\frac{\partial E}{\partial {\bf u}} & = & -\frac1\beta_X B{\bf x} ~.
\end{eqnarray}
and
\begin{eqnarray}
J \nabla_{\bf U} E  &=&   \left[ 
	\begin{array}{l}
		-\beta_X\beta_Y A \frac{\partial E}{\partial {\bf v}}\\
	-(-\beta_X\beta_Y A)^T \frac{\partial E}{\partial {\bf u}}
	\end{array}
\right] \nonumber\\
   &=&  \left[
	\begin{array}{l}
		-\beta_X A {\bf y}\\
		-\beta_Y B {\bf x}
	\end{array}
\right] 
	 =  \left[
	\begin{array}{l}
		\dot{\bf u}\\
		\dot{\bf v}
	\end{array}
\right] 
	 =  \dot{\bf U} ~. 
\label{HamiltonianForm}
\end{eqnarray}
We can transform ${\bf U}=({\bf u}, {\bf v})$ to canonical coordinates 
${\bf U}'=({\bf p}, {\bf q})$:
\begin{equation}
 \dot{\bf U}'=S\nabla_{{\bf U}'} E ~, 
\end{equation}
with
\begin{equation}
S = \left( \begin{array}{cc}
	O&-I\\
	I&O\\
	\end{array}
\right)
\end{equation}
where $I$ is an $N\times N$ identity matrix and with a linear transformation
${\bf U}'=M{\bf U}$ to the Hamiltonian form. \hfill $\Box$

\section{Reinforcement Schemes and Interaction Matrices}
\label{ReinforcementSchemesInteractionMatrices}

Here we give the reinforcement scheme interaction matrices for the
constant-environment collectives investigated in Sec. \ref{Sec:Examples}.

\subsection{Matching Pennies}

This game describes a non-transitive competition. Each agent chooses
a coin, which turns up either heads (H) or tails (T). Agent $X$
wins when the coins differ, otherwise agent $Y$ wins. Table
\ref{table:2mpgame} gives the reinforcement scheme for the various
possible plays. Note that the $\epsilon$s determine the size of
the winner's rewards. When $\epsilon_X+\epsilon_Y=0$, the game is
zero-sum. The Nash equilibrium is ${\bf x}^*={\bf y}^*=(1/2, 1/2)$.

Various extensions of Matching Pennies to more than two players are
known. We give the \emph{Even-Odd} game as an example for three agents 
$X$, $Y$, and $Z$ in a collective. All flip a coin. Agents $X$ and $Y$
win when the number of heads is even, otherwise $Z$ wins. Table
\ref{table:3mpgame} gives the reinforcement scheme. When the
$\epsilon$s add to zero, the game is zero-sum. The unique mixed Nash
equilibrium is ${\bf x}^*={\bf y}^*={\bf z}^* = (\frac12, \frac12,
\frac12)$---the simplex center. 

\begin{table}[htbp]
\vspace{5mm}
\centering
\begin{tabular}{@{}l@{}}
\noalign{\hrule height0.8pt}
\begin{tabular}{|cc|c|c|}
	\hline
	X&Y&$r^X$&$r^Y$\\
	\hline
	H&H&$-\epsilon_X$&$-\epsilon_Y$\\
	H&T&$\epsilon_{X}$&$\epsilon_{Y}$\\
	T&H&$\epsilon_{X}$&$\epsilon_{Y}$\\
	T&T&$-\epsilon_X$&$-\epsilon_Y$\\
	\hline
   \end{tabular}\\
   \noalign{\hrule height0.8pt}
\end{tabular}
\caption{The two-person Matching Pennies game: 
  $\epsilon_X\in(0.0,1.0]$ and $\epsilon_Y\in[-1.0,0.0)$.
  }
\label{table:2mpgame}
\end{table}

\begin{table}[htbp]
\centering
\begin{tabular}{@{}l@{}}
\noalign{\hrule height0.8pt}
\begin{tabular}{|ccc|c|c|c|}
\hline
X&Y&Z&$r^X$&$r^Y$&$r^Z$\\
\hline
H&H&H&$-\epsilon_X$&$-\epsilon_Y$&$-\epsilon_Z$\\
H&H&T&$\epsilon_{X}$&$\epsilon_{Y}$&$\epsilon_Z$\\
H&T&H&$\epsilon_{X}$&$\epsilon_{Y}$&$\epsilon_Z$\\
H&T&T&$-\epsilon_X$&$-\epsilon_Y$&$-\epsilon_Z$\\
T&H&H&$\epsilon_{X}$&$\epsilon_{Y}$&$\epsilon_Z$\\
T&H&T&$-\epsilon_X$&$-\epsilon_Y$&$-\epsilon_Z$\\
T&T&H&$-\epsilon_X$&$-\epsilon_Y$&$-\epsilon_Z$\\
T&T&T&$\epsilon_{X}$&$\epsilon_{Y}$&$\epsilon_Z$\\
\hline
\end{tabular}\\
\noalign{\hrule height0.8pt}
\end{tabular}
\caption{The three-player Even-Odd game: $\epsilon_X\in(0.0,1.0]$ and 
  $\epsilon_Y, \epsilon_Z\in[-1.0,0.0)$. 
  }
\label{table:3mpgame}
\end{table}

\newpage
\subsection{Rock-Scissors-Paper}

This game describes a non-transitive three-sided competition between
two agents: rock (R) beats scissors (S), scissors beats paper (P), but
paper beats rock. Table \ref{table:2rspgame} gives the reinforcement
scheme. The $\epsilon$s here control the rewards for ties.
When they add to zero, the game is zero-sum. The unique mixed Nash
equilibrium is ${\bf x}^*={\bf y}^*=(\frac13, \frac13, \frac13)$---again,
the center of the simplex. 

The extension of RSP interaction to three agents is straightforward.
The reinforcement scheme is given in Table \ref{table:3rspgame}. When
$\epsilon_X+\epsilon_Y+\epsilon_Z=0$, the game is zero-sum. The Nash
equilibrium is ${\bf x}^*={\bf y}^*={\bf z}^*=(1/3, 1/3, 1/3)$.

\begin{table}[htbp]
\centering
\begin{tabular}{@{}l@{}}
\noalign{\hrule height0.8pt}
\begin{tabular}{|cc|c|c|}
\hline
X&Y&$r^X$&$r^Y$\\
\hline
R&R&$\epsilon_X$&$\epsilon_Y$\\
R&S&1&-1\\
R&P&-1&1\\
S&R&-1&1\\
S&S&$\epsilon_X$&$\epsilon_Y$\\
S&P&1&-1\\
P&R&1&-1\\
P&S&-1&1\\
P&P&$\epsilon_X$&$\epsilon_Y$\\
\hline
\end{tabular}\\
\noalign{\hrule height0.8pt}
\end{tabular}
\caption{The two-person Rock-Scissors-Paper game: 
  ~$\epsilon_X, \epsilon_Y\in(-1.0,1.0)$. 
  }
\label{table:2rspgame}
\end{table}

\begin{table}[htbp]
\centering
  \begin{tabular}{@{}l@{}}
   \noalign{\hrule height0.8pt}
   \begin{tabular}{|ccc|c|c|c||ccc|c|c|c||ccc|c|c|c|}
	\hline
X&Y&Z&$r^X$&$r^Y$&$r^Z$ & X&Y&Z&$r^X$&$r^Y$&$r^Z$ & X&Y&Z&$r^X$&$r^Y$&$r^Z$\\
\hline
R&R&R&$\epsilon_{X}$&$\epsilon_{Y}$&$\epsilon_{Z}$ & S&R&R&-2&1&1 & P&R&R&2&-1&-1\\
R&R&S&1&1&-2 & S&R&S&-1&2&-1 & P&R&S&$\epsilon_{X}$&$\epsilon_{Y}$&$\epsilon_{Z}$\\
R&R&P&-1&-1&2 & S&R&P&$\epsilon_{X}$&$\epsilon_{Y}$&$\epsilon_{Z}$ & P&R&P&1&-2&1\\
R&S&R&1&-2&1 & S&S&R&-1&-1&2 & P&S&R&$\epsilon_{X}$&$\epsilon_{Y}$&$\epsilon_{Z}$\\
R&S&S&2&-1&-1 & S&S&S&$\epsilon_{X}$&$\epsilon_{Y}$&$\epsilon_{Z}$ & P&S&S&-2&1&1\\
R&S&P&$\epsilon_{X}$&$\epsilon_{Y}$&$\epsilon_{Z}$ & S&S&P&1&1&-2 & P&S&P&-1&2&-1\\
R&P&R&-1&2&-1 & S&P&R&$\epsilon_{X}$&$\epsilon_{Y}$&$\epsilon_{Z}$ & P&P&R&1&1&-2\\
R&P&S&$\epsilon_{X}$&$\epsilon_{Y}$&$\epsilon_{Z}$ & S&P&S&1&-2&1 & P&P&S&-1&-1&2\\
R&P&P&-2&1&1 & S&P&P&2&-1&-1 & P&P&P&$\epsilon_{X}$&$\epsilon_{Y}$&$\epsilon_{Z}$\\
	\hline
   \end{tabular}\\
   \noalign{\hrule height0.8pt}
  \end{tabular} 
\caption{The 3-person Rock-Scissors-Paper game: 
  ~$\epsilon_X, \epsilon_Y, \epsilon_Z\in(-1.0,1.0)$. 
  }
\label{table:3rspgame}
\end{table}

\end{appendix}

\bibliography{SDCA_pre}

\end{document}